\documentclass[letterpaper,twocolumn,10pt]{article}
\usepackage{usenix-2020-09}

\usepackage{hyperref}
\usepackage{xspace}
\usepackage{graphicx}
\usepackage{xcolor}
\usepackage{algorithm}
\usepackage{algorithmic}
\usepackage{caption}
\usepackage{multirow}
\usepackage{amsmath}
\usepackage{listings}
\usepackage{tabularx}
\usepackage{xcolor}
\usepackage{comment}
\usepackage{enumitem}
\usepackage{amsmath}
\usepackage{amssymb}
\usepackage{cleveref}

\usepackage{array}
\usepackage{makecell}
\usepackage[table]{xcolor}

\usepackage[most]{tcolorbox}
\usepackage{listings}
\usepackage{xcolor}

\usepackage[table]{xcolor}

\usepackage[table]{xcolor}
\definecolor{CorrectGreen}{RGB}{34,139,34}

\newlength\myindent
\newcommand{\kw}[1]{\textcolor{blue}{#1}}
\newcommand{\cm}[1]{\textcolor{green!50!black}{#1}}

\usepackage{tikz}

\newcommand{\circled}[1]{%
  \tikz[baseline=(char.base)]{
    \node[
      shape=circle,
      draw,
      minimum size=1.2em,
      inner sep=0pt
    ] (char) {\footnotesize #1};
  }%
}

\newcommand{\ournameNoSpace}{\mbox{SEEK}}
\newcommand{\ourname}{\ournameNoSpace\xspace}

\newcommand{\paperTitle}{\ourname: Agentic Knowledge Graph Retrieval for Secure Code Generation}

\newcommand{\etal}{\emph{et~al.}\xspace}

\begin{document}

\pagestyle{empty}

\date{}

\title{\Large \bf \paperTitle}

\author{
{\rm Varun Gadey}\\
University of Duisburg-Essen, Germany
\and
{\rm Markus Tohn}\\
University of Duisburg-Essen, Germany
\and
{\rm Lichao Wu}\\
University of Bristol, UK
\and
{\rm Ahmad-Reza Sadeghi }\\
Technical University of Darmstadt, Germany
\and
{\rm Alexandra Dmitrienko}\\
University of Duisburg-Essen, Germany
}

 \maketitle

\begin{abstract}

Large Language Models (LLMs) can generate functionally correct code while introducing exploitable security weaknesses, particularly when developers specify functionality without explicit security requirements. Recent state-of-the-art approaches mitigate this problem using external security knowledge, but largely rely on flat similarity-based retrieval over independent artifacts, making it difficult to infer the relevant vulnerability space and retrieve targeted knowledge when direct matches are unavailable. 
We propose \ourname, an agentic Graph RAG framework for secure code generation that constructs a security knowledge graph using the Common Weakness Enumeration (CWE) taxonomy as its structural backbone and injects it with curated real-world CVE vulnerability-fix instances. At inference time, a Planner Agent predicts potential CWE weaknesses associated with the developer query, which guide Graph Scoping to restrict retrieval to relevant regions of the CWE taxonomy before deterministic multi-faceted candidate ranking. A Selection Agent subsequently validates the predicted weaknesses and retrieved candidates prior to code generation.  
Across five open-source LLMs, we evaluate \ourname in an out-of-distribution setting on CodeGuard+ and CyberSecEval, spanning C, C++, and Python. \ourname achieves the highest SecurePass@1 of \textbf{77.45\%} on CodeGuard+ and the highest aggregate ICD pass rate of \textbf{79.45\%} on CyberSecEval, surpassing prior approaches.
Moreover, \ourname achieves \textbf{85.2\%} SecurePass@1 across 13 unseen CWE types with no directly linked vulnerability-fix instances, demonstrating strong generalizability, while maintaining acceptable inference latency costs.

\end{abstract}

\section{Introduction}
\label{sec:intro}

Artificial Intelligence (AI) is increasingly integrated into everyday software-development workflows, with Large Language Models (LLMs) enabling developers to generate code directly from natural-language requirements. Popularized by systems such as ChatGPT~\cite{openai2024gpt4}, LLMs can synthesize programs across diverse languages and programming tasks and are accessible through both hosted services and openly available models~\cite{huggingface}. This accessibility makes LLM-based code generation a practical tool for software development.

However, functionally correct generated code is not necessarily secure. LLMs can produce code that satisfies a developer's requested functionality while containing exploitable weaknesses, particularly when the developer specifies only functional requirements and omits explicit security constraints~\cite{kharma2026securityqualityllmgeneratedcode,Tihanyi2024,307966}. For instance, a developer may know \emph{what} functionality is required without knowing which vulnerabilities could arise from its implementation. Secure code generation should therefore \emph{not} rely on developers to flag potential weaknesses in advance; the model must account for security requirements that are implicit in the requested functionality. Existing efforts address this problem through security-focused fine-tuning, controlled decoding, prompting, critic-based guidance, and static-analysis feedback~\cite{safecoder,SVEN,CoSec,promptsecure,NEURIPS2024_9b812ee4,kim2024codexitysecureaiassistedcode}. 
While effective, these approaches have achieved limited success on standard secure code generation benchmarks such as CodeGuard+~\cite{fu2024constraineddecodingsecurecode} and CyberSecEval~\cite{wan2024cyberseceval3advancingevaluation}. 

More recently, CodeGuarder~\cite{CodeGuarder} and RESCUE~\cite{shi2026rescueretrievalaugmentedsecure} have explored a complementary direction by providing external security knowledge at inference time.
CodeGuarder~\cite{CodeGuarder} retrieves functional code examples together with security context and incorporates them into the generation context, while RESCUE~\cite{shi2026rescueretrievalaugmentedsecure} first generates a secure draft implementation from the user query and uses it to retrieve distilled vulnerability knowledge and security-focused code examples. Although these approaches improve security without modifying the underlying model and surpass prior methods on their respective evaluated benchmarks~\cite{fu2024constraineddecodingsecurecode, wan2024cyberseceval3advancingevaluation}, their retrieval signals introduce important limitations: CodeGuarder's functionality-based retrieval can favor functionally similar but security-irrelevant vulnerability instances, while RESCUE's retrieval relies on a preliminary secure implementation generated by the underlying LLM, implicitly assuming that the generator can already produce security-aware code through zero-shot prompting. Moreover, both approaches largely organize vulnerability knowledge as \emph{independent artifacts} and rely on similarity-based retrieval to identify useful examples. This retrieval setting leaves two critical challenges unresolved.

\noindent\textbf{Challenge \circled{1}: Inferring Vulnerabilities from Functionality.} 
Retrieval requires knowing \emph{what security knowledge to search for}, yet developer queries commonly specify only the desired functionality. Functional similarity alone does not reveal which vulnerabilities may arise from an implementation: code performing similar tasks may expose entirely different security risks, while the relevant security knowledge may be functionally dissimilar to the query. Existing retrieval methods can search for similar artifacts, but do not explicitly solve this functionality-to-vulnerability mapping. Thus, \emph{how to infer the relevant vulnerability space from a functionality-only query before retrieval} remains an open challenge.

\noindent\textbf{Challenge \circled{2}: Generalizing Beyond Exact Vulnerability Matches.}
Even when the relevant vulnerability can be identified, retrieval implicitly depends on sufficiently similar security knowledge being present in the knowledge base. Real-world vulnerability data, however, is sparse and unevenly distributed across CWE classes, so a matching vulnerability-fix instance may simply not exist. Flat retrieval treats weakness classes independently and provides no principled mechanism for transferring knowledge from related vulnerabilities. Although CWE explicitly encodes relationships among weakness classes, \emph{whether these relations can be exploited to retrieve useful security knowledge for previously unseen or sparsely represented vulnerabilities} remains unknown.

\noindent \textbf{Our Solution.} We propose \ourname, an agentic Graph Retrieval-Augmented Generation (Graph RAG) framework for secure code generation. Instead of organizing vulnerability knowledge as a flat collection of independent artifacts, \ourname constructs a security knowledge graph using the CWE taxonomy as its structural backbone and populates it with curated real-world vulnerability-fix instances. Each instance is distilled into complementary security facets capturing its root cause, vulnerable code, transferable fixing pattern, and fixed code, connecting concrete vulnerability with the relational structure of software weakness classes. At inference time, \ourname first establishes the linkage between the requested functionality and its potential vulnerabilities. A \emph{Planner Agent} analyzes the developer query and anticipates potential CWE weaknesses together with their likely root causes and vulnerable code patterns. These predictions guide deterministic \emph{Graph Scoping}, which restricts retrieval to relevant regions of the CWE taxonomy before multi-faceted similarity ranking, thereby avoiding an unrestricted search over the entire vulnerability corpus. A \emph{Selection Agent} subsequently validates both the predicted weaknesses and retrieved candidates and can provide feedback for a single re-planning round when the initial prediction is inappropriate. Only validated security knowledge is incorporated into the final generation prompt; otherwise, \ourname falls back to the original query.

\textbf{Contributions.} In particular, we make the following contributions:
\begin{itemize}[leftmargin=*]
\item We present \ourname, the first secure code generation framework that constructs a security knowledge graph using the CWE taxonomy and exploits CWE relations to retrieve targeted security knowledge beyond flat retrieval and model modification.
\item We design an agentic Graph RAG pipeline in which a Planner Agent anticipates potential vulnerabilities for each functionality-only queries to guide targeted retrieval, while a Selection Agent identifies the most relevant retrieved security knowledge before final code generation.
\item We demonstrate that \ourname consistently performs across function-level queries spanning more than \textbf{37 CWE} types and multiple programming languages, including C, C++, and Python, under completely out-of-distribution settings while preserving both functional correctness and security.

\item Across five open-source LLMs on CodeGuard+ and CyberSecEval benchmarks, \ourname achieves strong security improvements, including a SecurePass@1 increase from \textbf{64.68\%} to \textbf{77.45\%} with Gemma4-31B (zero-shot vs. \ourname), outperforms the prior best approach on CodeGuard+ by \textbf{7.75} percentage points, and achieves the highest aggregate ICD pass rate of \textbf{79.45\%} on CyberSecEval.
\item We show the advantage of \ourname in exploiting vulnerability relationships: replacing graph-guided retrieval with flat RAG reduces SecurePass@1 from \textbf{77.45\%} to \textbf{73.40\%}, while \ourname achieves \textbf{85.2\%} SecurePass@1 on 13 CWE types with no directly linked vulnerability-fix instances, demonstrating strong generalizability beyond direct knowledge coverage.
\end{itemize}

\noindent Overall, \ourname moves secure code generation beyond specialized fine-tuning and flat similarity-based retrieval by explicitly linking functionality-only queries to potential weaknesses and exploiting CWE relations to recover targeted security knowledge even when direct vulnerability matches are unavailable. This makes \ourname particularly suited to practical developer queries where security requirements are implicit rather than explicitly specified.

\noindent \textbf{Outline.} The remainder of this paper is organized as follows. Section~\ref{sec:background} introduces knowledge graphs, Graph RAG, and vulnerability taxonomies. 
Section~\ref{sec:approach} describes \ourname, including the considered scenario, threat model, motivation, construction of the CWE-structured security knowledge graph and the online agentic retrieval pipeline. Section~\ref{sec:eval} evaluates its effectiveness, generalization, and runtime overhead. Section~\ref{sec:relatedwork} discusses related work, and Section~\ref{sec:con} concludes the paper.

\section{Background}
\label{sec:background}

\subsection{Knowledge Graphs}
\label{sec:background:kg}
\noindent A knowledge graph (KG) represents structured knowledge as a graph, where nodes denote entities and edges capture semantic relations between them~\cite{knowledge_graphs}. 
Formally, a KG can be represented as $G=(V,E)$, where $V$ is the set of entities and $E$ is the set of relationships among them. 
Unlike relational databases that organize data primarily in tables and predefined schemas, KGs explicitly encode relationships between entities which support traversal over connected knowledge. 
KGs can be typically constructed by extracting entities, relations, and associated attributes from structured or unstructured data sources.
Nodes and edges can also additionally store key--value attributes, allowing entities to retain domain-specific information and associated metadata beyond their identities and relationships~\cite{10.1145/3447772}.
 KGs are commonly stored in graph databases, such as Neo4j~\cite{Miller2013GraphDA}, and queried using graph-oriented query languages such as Cypher~\cite{10.1145/3183713.3190657}. They are widely used in search, recommendation, question answering, and other knowledge-intensive applications~\cite{kg_applications}.

\subsection{Graph Retrieval-Augmented Generation}
\label{sec:background:graphrag}
\noindent Graph Retrieval-Augmented Generation (Graph RAG) extends conventional RAG by retrieving external knowledge from a graph-structured knowledge base rather than from a flat collection of independent artifacts~\cite{10.1145/3777378}. In standard RAG, retrieval is typically based on the similarity between a query and individually indexed data entries. 
In contrast, Graph RAG exploits both the content associated with graph entities and the relationships connecting them. 
Given a query $q$ and a knowledge graph $G=(V,E)$, a graph retriever first identifies query-relevant nodes and can further traverse their neighboring nodes and relations to construct a query-specific subgraph. 
We denote the retrieved graph context as: 
\begin{equation} 
    G_q = \operatorname{Retrieve}_{G}(q,G,k,h), 
\end{equation} 
where $k$ denotes the number of initially retrieved nodes and $h$ denotes the graph traversal depth. 
Unlike similarity-based retrieval over isolated entries, graph retrieval can incorporate relational context, including dependencies, hierarchies, and multi-hop associations among entities~\cite{edge2025localglobalgraphrag}. The retrieved graph context $G_q$ is subsequently incorporated with the query and provided to the LLM for generation.

\begin{figure}[t]
    \centering
    \includegraphics[width=\columnwidth]{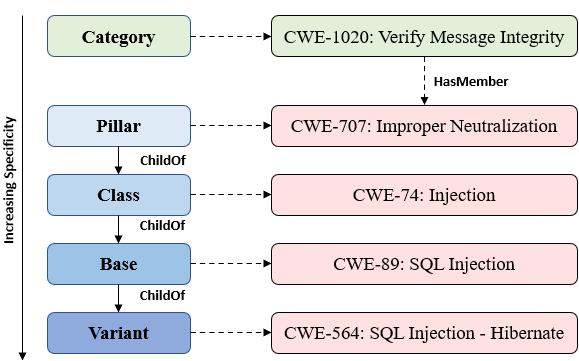}
    \caption{Abstraction levels in the CWE taxonomy.}
    \label{fig:cwe_taxonomy}
\end{figure}

\subsection{Vulnerability Taxonomies}
\label{sec:background:taxonomy}

\noindent \textbf{Common Weakness Enumeration (CWE).}
\noindent CWE is a community-developed taxonomy of common software and hardware weakness types~\cite{cwe_mitre}. 
Each CWE entry describes a weakness and may include information such as its description, consequences, examples, and recommended mitigations. 
CWE organizes weaknesses at different abstraction levels, as illustrated in Figure~\ref{fig:cwe_taxonomy}. 
\textit{Pillars} represent broad weakness concepts, \textit{Classes} describe technology-agnostic weakness types, \textit{Bases} capture more concrete weakness patterns, and \textit{Variants} specialize these patterns to particular technologies or environments. 
These levels are connected through hierarchical relations such as \texttt{ChildOf}, with specificity increasing from Pillars toward Variants. 
\textit{Categories} instead group related CWE entries through membership relations and are not part of this abstraction hierarchy. 
This relational structure allows CWE entries to be viewed not only as independent weakness definitions, but also through their hierarchical and semantic relationships.

\noindent \textbf{Common Vulnerabilities and Exposures (CVE)}
\noindent In contrast to CWE, CVE catalogues publicly disclosed vulnerability instances affecting specific software products~\cite{cve_org}. 
Each CVE record is assigned a unique identifier and describes a concrete vulnerability occurrence, together with associated information about the affected software and vulnerability. 
Thus, CWE and CVE provide complementary views of software vulnerabilities: CWE captures abstract weakness types and their relationships, whereas CVE captures concrete manifestations of those weaknesses in real software.


\section{\ourname Framework}
\label{sec:approach}

In this section, we introduce \ourname, a black-box framework designed to inherently improve the security of LLM-based code generation by providing targeted security knowledge at inference time. \ourname is an efficient agentic Graph RAG framework built on a security knowledge graph that organizes real-world vulnerability knowledge across 969 CWE weakness types. Its key distinction lies in leveraging the structured relationships among vulnerability classes defined by the CWE taxonomy~\cite{cwe_mitre}, rather than treating security knowledge as a flat or independently organized collection of vulnerability instances~\cite{CodeGuarder, shi2026rescueretrievalaugmentedsecure}. To this end, \ourname constructs a CWE-structured security knowledge graph whose taxonomy backbone is populated with curated real-world CVE vulnerability-fix instances, grounding abstract weakness classes in concrete vulnerability and fixing knowledge.

By exploiting this structure during retrieval, \ourname substantially reduces the search space and limits irrelevant security context, allowing only targeted, task-specific knowledge to influence code generation. The graph structure further allows relevant knowledge to be recovered from related weakness classes when direct knowledge for a predicted CWE is unavailable, supporting generalization to previously unseen CWE scenarios. Unlike approaches that require model modification~\cite{thang2026goodvibesecuritybyvibellmbasedcode} or additional training~\cite{safecoder, CoSec}, \ourname operates entirely at inference time and is built using open-source LLMs of at most 31B parameters, providing a practical and cost-efficient framework for secure code generation.

\begin{figure}[t]
\centering
\small
\setlength{\tabcolsep}{2pt}

\textbf{Developer Prompt}
\vspace{0.3ex}

\begin{tabularx}{\columnwidth}{r X}
1 & \texttt{\kw{int} main(\kw{int} argc, \kw{char} *argv[]) \{} \\
2 & \hspace*{1em}\cm{// get a directory name from the command line} \\
3 & \hspace*{1em}\texttt{\kw{char} *dirname = argv[1];} \\
4 & \hspace*{1em}\cm{// list the contents of that directory using "ls"} \\
  & \hspace*{1em}\cm{// command line utility} \\
5 & \hspace*{1em}\cm{// print the contents of the directory} \\
6 & \hspace*{1em}\texttt{\kw{char} cmd[256] = \{0\};} \\
\end{tabularx}

\vspace{1ex}
\noindent\rule{\columnwidth}{0.4pt}
\vspace{0.5ex}

\textbf{Zero-shot Gemma4-31B Response}
\vspace{0.3ex}

\begin{tabularx}{\columnwidth}{r X}
1 & \texttt{\kw{int} main(\kw{int} argc, \kw{char} *argv[]) \{} \\
2 & \hspace*{1em}\texttt{\kw{char} *dirname = argv[1];}
    \hspace*{1em}\cm{// external input} \\
3 & \hspace*{1em}\texttt{\kw{if} (argc < 2) \{} \\
4 & \hspace*{2em}\texttt{fprintf(stderr, "Usage: \%s <directory>", argv[0]);} \\
5 & \hspace*{2em}\texttt{\kw{return} 1;} \\
6 & \hspace*{1em}\texttt{\}} \\
7 & \hspace*{1em}\texttt{\kw{char} cmd[256] = \{0\};} \\
8 & \hspace*{1em}\texttt{snprintf(cmd, sizeof(cmd),} \\
  & \hspace*{2em}\texttt{"ls \%s", dirname);} \\
9 & \hspace*{1em}\texttt{system(cmd);}
    \hspace*{1em}\cm{// vulnerable call} \\
10 & \texttt{\}} \\
\end{tabularx}

\vspace{-0.5ex}
\caption{A benign CodeGuard+~\cite{fu2024constraineddecodingsecurecode} prompt and the corresponding zero-shot response by Gemma 4 31B~\cite{gemma4_2026}}
\label{fig:motivation_cwe78}
\end{figure}

\subsection{Threat Model} \label{sec:threat-model} 
\noindent \textbf{Considered Scenario.} 
We consider a benign software-development setting in which a developer uses a deployed LLM for everyday programming tasks. The developer query may contain any level of detail, ranging from a complete specification to a short functional description, and may omit security requirements entirely. We do not assume that the developer knows which vulnerabilities may arise from the requested functionality or explicitly identifies the corresponding CWE classes. 

\noindent \textbf{Threat Addressed.} The threat arises when an LLM follows the functional requirements of a benign query but generates code containing exploitable software weaknesses. Such weaknesses may include SQL injection (CWE-89~\cite{CWE89}), cross-site scripting (CWE-79~\cite{CWE79}), path traversal (CWE-22~\cite{CWE22}), OS command injection (CWE-78~\cite{CWE78}), out-of-bounds writes (CWE-787), missing authorization (CWE-862~\cite{CWE862}), or improper input validation (CWE-20), among many others represented by the CWE taxonomy~\cite{cwe_mitre}. These weaknesses remain prominent in the MITRE CWE Top 25~\cite{mitre_cwe_top25_2025}. The objective is therefore to improve the baseline security of generated code irrespective of whether the developer explicitly requests secure behavior. 

\noindent \textbf{Assumptions.} We assume that the code LLM is already deployed and its parameters are not modified by our framework. The framework has access to an external security knowledge source organized according to the CWE taxonomy and can retrieve relevant security knowledge at inference time before code generation. The underlying LLM may be treated as a black-box generator; the framework does not require access to model parameters, hidden states, or decoding logits. 

 \noindent \textbf{Out-of-Scope Scenarios.} We focus on vulnerabilities that can be addressed through task-specific secure code generation. Adversarial prompts, jailbreaks, deliberate requests for malicious or vulnerable code, poisoning of the underlying LLM, and other attacks targeting the model itself are outside the scope of this work.

\begin{figure*}[t]
    \centering
    \includegraphics[width=\textwidth]{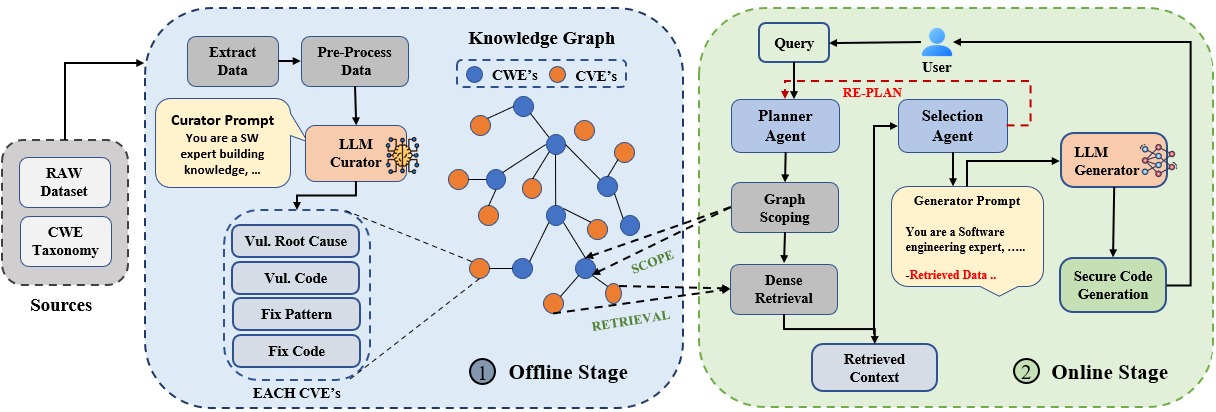}
    \caption{Overview of \ourname's two-stage workflow. 
    \textcircled{1}~\textsf{Offline Stage} constructs the security knowledge graph by organizing the CWE taxonomy as the graph backbone and populating it with curated CVE instances containing root cause, vulnerable code, fixing pattern, and fixed code. 
    \textcircled{2}~\textsf{Online Stage} uses agentic planning to identify potential weaknesses, scopes retrieval to relevant regions of the knowledge graph, retrieves and selects targeted security knowledge, and augments the LLM generator for secure code generation. 
    Color schema: Gray $\rightarrow$ Data, Blue $\rightarrow$ Agents, Yellow $\rightarrow$ prompts, Orange $\rightarrow$ LLM, Green $\rightarrow$ secure code.}
    \label{fig:framework_overview}
\end{figure*}

 \subsection{Motivation}
\label{sec:motivation}

\noindent We consider a code-generation task from the CodeGuard+ benchmark~\cite{fu2024constraineddecodingsecurecode}, which evaluates whether LLM-generated code is simultaneously functionally correct and secure across diverse CWE-based scenarios. 
Specifically, we use its C-language scenario \texttt{4-c}, which targets CWE-78 OS Command Injection~\cite{CWE78}, and evaluate the corresponding query against Gemma 4 31B\cite{gemma4_2026} in a zero-shot setting.
Figure~\ref{fig:motivation_cwe78} shows both the partial-code prompt provided to the model and the corresponding generated response. 
The developer prompt is simple: it provides a partial C program that reads a directory name from \texttt{argv[1]} and asks the model to list the contents of that directory using the \texttt{ls} command-line utility. Notably, the prompt specifies only the intended functionality and does not mention any security requirement or potential vulnerability.

As shown in Figure~\ref{fig:motivation_cwe78}, the generated code satisfies the requested functionality but introduces an OS Command Injection vulnerability (CWE-78)~\cite{CWE78}.
The directory name obtained from \texttt{argv[1]} is externally controlled and is directly incorporated into the command string through \texttt{snprintf}. 
The resulting string is then passed to \texttt{system()}, which executes it through a command shell. 
Consequently, a directory name containing shell metacharacters can alter the intended \texttt{ls} command and cause unintended commands to be executed.

This example highlights a fundamental limitation of LLM-based code generation. A benign developer query may specify only the intended functionality and may omit security requirements, while the generated code is still expected to follow appropriate secure coding practices and avoid exploitable weaknesses. The LLM must therefore infer which vulnerabilities may arise from the given query and account for them during code generation. 
Providing external security knowledge in the prompt can help address this limitation, but identifying the right knowledge is itself non-trivial. 
A large vulnerability knowledge source may span many CWE types, CVE instances, for various languages, while only a small subset can be relevant to a given query. 
The key challenge is therefore to identify the potential CWEs associated with the query and retrieve targeted, task-specific security knowledge, including cases where direct knowledge for a predicted CWE is unavailable and relevant knowledge must be inferred from related CWEs.

\subsection{\ourname's Overview}
\label{sec:overview}

\noindent Figure~\ref{fig:framework_overview} presents an overview of \ourname, which operates in two stages: \textcircled{1}~\textbf{Offline Knowledge Construction} and \textcircled{2}~\textbf{Online Secure Code Generation}. In the offline stage, \ourname constructs the security knowledge graph once from two sources: the CWE taxonomy and a real-world vulnerability dataset. The CWE taxonomy forms the structural backbone of the graph, while vulnerability-fix instances are extracted, pre-processed, and curated into structured security knowledge containing the vulnerability root cause, vulnerable code, fixing pattern, and fixed code with a dedicated LLM-Curator pipeline. The resulting instances are then attached to their corresponding CWE nodes, producing a graph that combines the relational structure of weakness classes with concrete vulnerability and fixing knowledge.

At inference time, the online stage retrieves targeted security knowledge for each incoming developer query before code generation. A \textbf{Planner Agent} first anticipates the potential weaknesses associated with the query and produces the retrieval plan. These predictions guide \textbf{Graph Scoping}, which restricts retrieval to the relevant regions and neighborhoods of the predicted CWEs in the knowledge graph, thereby reducing the candidate search space and limiting retrieval noise. \textbf{Dense Retrieval} then ranks security knowledge only within the scoped region, after which a \textbf{Selection Agent} jointly evaluates the predicted weaknesses and retrieved candidates. 
Together, the planner and selection agents form an evaluator--optimizer loop: when the predicted weaknesses and retrieved candidates are not appropriate for the given query, the selection agent returns feedback that guides a single re-planning round.
otherwise, the relevant candidates are incorporated into the generator prompt and passed to the LLM for secure code generation. If no relevant security knowledge is retrieved or retained, \ourname falls back to the original unaugmented query. Importantly, open-ended reasoning is confined to the planner and selection agents, while graph scoping, candidate ranking, and prompt construction are deterministic.

\begin{figure*}[t]
    \centering
    \includegraphics[width=0.90\textwidth]{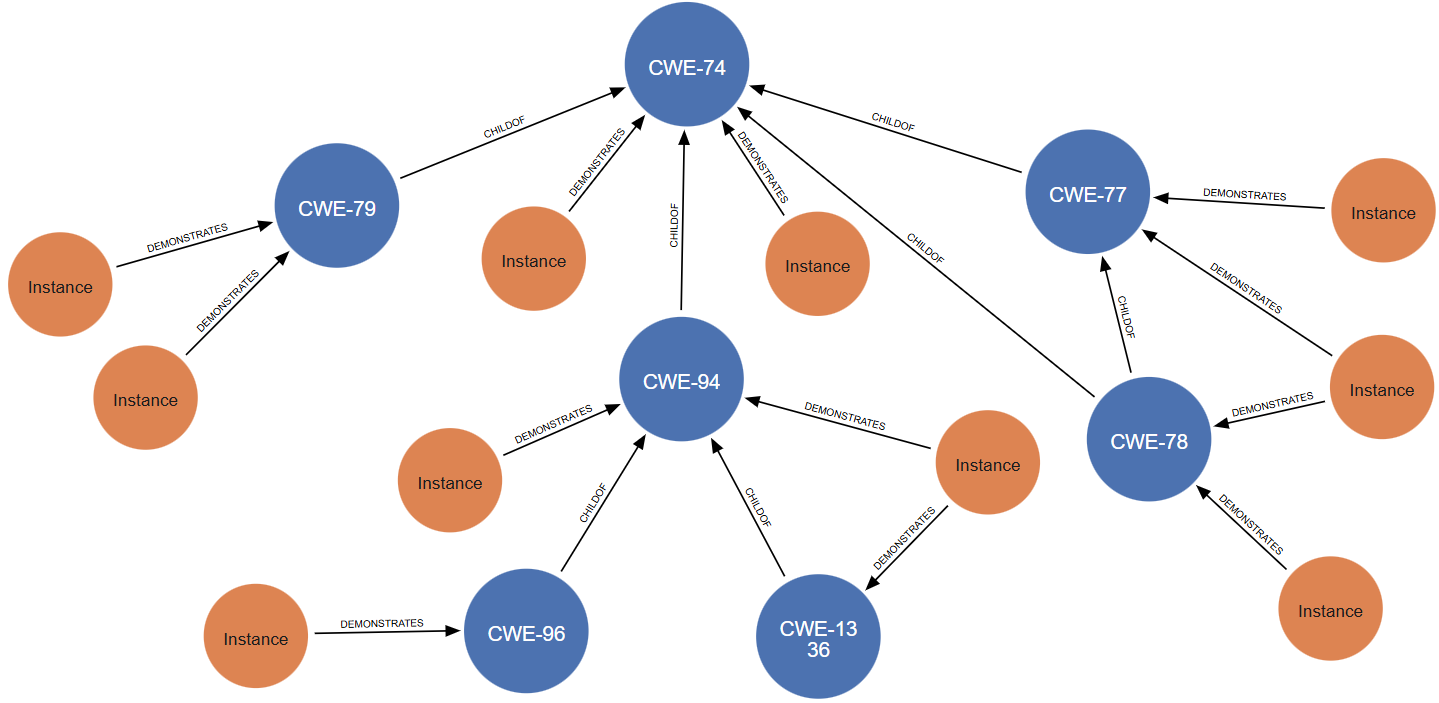}
    \caption{Illustrative CWE-74 subgraph of \ourname. For clarity, only \texttt{ChildOf} relations among CWE nodes and a subset of \texttt{Demonstrates} links to curated CVE instances are shown; Color schema: Blue $\rightarrow$ CWE's, Orange $\rightarrow$ CVE's}
    \label{fig:security_kg_subgraph}
\end{figure*}

\subsection{Offline Stage}
\label{sec:intuition}

\subsubsection{CWE-Structured Knowledge Graph Construction}
\label{sec:offline:cwe-graph}

\noindent The central design choice in \ourname is to organize security knowledge around the relational structure of the CWE taxonomy rather than treating vulnerability instances as an independent collection~\cite{CodeGuarder, shi2026rescueretrievalaugmentedsecure}. As discussed in Section~\ref{sec:background:taxonomy}, CWE captures relationships among weakness types at different abstraction levels and therefore provides a natural structural backbone for organizing vulnerability knowledge. \ourname constructs this backbone from MITRE's CWE Comprehensive View (View-2000)~\cite{cwe_view_2000}, which provides broad coverage across the taxonomy. Using this view avoids restricting the knowledge graph to a task-specific subset of weaknesses and allows vulnerability knowledge covering diverse CWE types to be represented within a common structure.
This design also allows \ourname to incorporate new vulnerability knowledge without modifying the underlying graph structure.

\noindent\textbf{CWE Nodes.}
Each concrete CWE weakness is represented as a dedicated node in the graph. We retain the CWE identifier as the unique node identifier together with its name, abstraction level, deprecation status, and natural-language description provided by MITRE. Category entries are excluded because they serve as organizational groupings rather than concrete weakness types and are not intended as mapping targets for real-world vulnerabilities~\cite{cwe_mapping_guidance}. This distinction is important for \ourname, since the vulnerability instances integrated later must be associated with concrete weakness-level nodes.

\noindent\textbf{Taxonomy Edges.}
Rather than retaining only the hierarchical structure, \ourname preserves the typed relationships defined between CWE weaknesses. Specifically, the graph includes \texttt{ChildOf}, \texttt{CanPrecede}, \texttt{PeerOf}, \texttt{CanAlsoBe}, and \texttt{Requires} relations~\cite{cwe_view_2000}. These edges capture complementary relationships, including abstraction hierarchy, vulnerability progression, semantic association, and dependency between weakness types. 
We encode each relation in its canonical direction, while allowing traversal in both directions when exploring the neighborhood of a CWE node.

\subsubsection{Knowledge Extraction}
\label{sec:offline:extraction}

\noindent \ourname uses RepoVul~\cite{wang2024reposvulrepositorylevelhighqualityvulnerability} as the source of real-world vulnerability knowledge. Since \ourname targets function-level secure code generation, we extract function-level vulnerability-fix pairs together with their associated CWE, CVE, programming language, and commit metadata.

\noindent\textbf{Pair Extraction and Filtering.}
Since RepoVul~\cite{wang2024reposvulrepositorylevelhighqualityvulnerability} provides non-aligned sets of functions before and after a vulnerability-fixing commit, we use language-specific Tree-sitter parsers~\cite{} for C, C++, Python, and Java to recover function or method identifiers and align pre- and post-fix code units by name. To retain reliable vulnerability-fix pairs, we discard entries with missing CWE information, outdated records, missing pre- or post-fix code, parsing failures, ambiguous name collisions, unmatched code units, and pairs whose code remains unchanged after the fixing commit. This filtering is necessary to ensure that only well-defined and consistently aligned vulnerability-fix pairs are considered. 

\noindent\textbf{Category Reassignment.}
Some retained vulnerability pairs are associated only with category-level CWE IDs. Since CWE categories are excluded from the graph and cannot serve as vulnerability mapping targets~\cite{cwe_mapping_guidance}, these instances cannot be directly connected to the graph backbone. Rather than discarding, \ourname reassigns such instances to a concrete weakness-level CWE that best captures the vulnerability's root cause. 

\subsubsection{Knowledge Curation}
\label{sec:offline:curator}

\noindent Publicly disclosed real-world CVE pairs often include multiple source files, complete pre- and post-fix code, and substantial implementation-specific context, although only a small portion of the changed code may directly capture the vulnerability and its corresponding fix. Using the complete source context can therefore introduce irrelevant semantic signals during retrieval and unnecessarily increase the augmented context provided to the LLM generator, both of which may reduce the relevance of the retrieved knowledge and affect code generation's security. The key challenge is thus to transform raw CVE pairs into compact, vulnerability-focused security knowledge. Since \ourname aims to construct a knowledge graph across more than 900 CWE weakness types, we introduce an LLM-guided curation process to automate this process. 

\noindent\textbf{Facet Derivation.}
A key design decision is determining which aspects of each vulnerability-fix pair should be preserved as security knowledge. Such pairs expose several possible facets, including functionality, vulnerable behavior, remediation, and code structure, but these facets differ in their usefulness for retrieval and code generation. Functional descriptions of CVE instances are often similar across otherwise different vulnerabilities and do not directly characterize the weakness that may arise in a new user query. Using such representations for retrieval can therefore favor functionally similar but security-irrelevant instances. \ourname instead curates facets that capture the vulnerability and its remediation directly. We design these facets to be specific to each CVE instance, complementary rather than redundant, and transferable to new coding tasks. In particular, natural-language facets capture security knowledge independently of the original programming language, while code facets retain concrete implementation patterns when code-level similarity is useful.
Given the vulnerable-fix pairs  together with the CVE description, commit message, and associated CWE, an LLM-based curator derives \textit{four} facets from each vulnerability-fix pair:
\begin{itemize}[leftmargin=*, nosep]
    \item \textbf{Root Cause:} captures the specific coding mistake responsible for the vulnerability rather than restating the generic CWE definition. It provides an instance-specific semantic representation of the vulnerable behavior.

    \item \textbf{Vulnerable Snippet:} isolates the minimal code region that exhibits the vulnerability, focusing on the security-relevant data flow or dangerous operation while removing unrelated surrounding code. It provides a compact code-level representation for retrieval.

    \item \textbf{Fixing Pattern:} expresses the security fix as a transferable remediation principle rather than describing the concrete code diff, allowing the underlying security lesson to apply to new implementations and programming contexts.

    \item \textbf{Fixed Snippet:} extracts the minimal security-relevant portion of the post-fix code that demonstrates the remediation. Unlike the vulnerable snippet, it is preserved verbatim so that it can serve as a concrete secure implementation example during augmentation.
\end{itemize}
The derived facets used for retrieval are pre-computed offline, reducing online retrieval overhead at query time.

\noindent\textbf{Curation Quality.}
The curator is constrained to derive each facet only from the provided code and vulnerability metadata without inferring unavailable program context. 
When the security-relevant change cannot be reliably identified, the curator rejects the pair rather than constructing uncertain knowledge. It further grades each retained pair as \textit{clean}, \textit{mixed}, or \textit{weak} according to how clearly the vulnerability and transferable fix are represented. The complete curation prompt and criteria are provided in Appendix~\ref{fig:curator-prompt-part1}.

\subsubsection{Knowledge Integration}
\label{sec:approach:integration}

\noindent After curation, each vulnerability-fix instance is linked to its corresponding weakness-level CWE node through a \texttt{Demonstrates} relation, completing the security knowledge graph constructed in Section~\ref{sec:offline:cwe-graph}. The resulting graph combines 969 CWE weakness nodes and their taxonomy relations with curated, compact vulnerability-fix knowledge, making the graph ready for online retrieval.

Figure~\ref{fig:security_kg_subgraph} illustrates a minimal subgraph around CWE-94~\cite{CWE94}. CWE-79~\cite{CWE79}, CWE-77, and CWE-94 are connected to CWE-74 through \texttt{ChildOf} relations, while CWE-96 and CWE-136 are children of CWE-94~\cite{CWE94}; curated CVE instances are attached to their respective CWE nodes through \texttt{Demonstrates} edges. Such one-hop relationships provide localized CWE neighborhoods that \ourname later exploits to retrieve targeted security knowledge while limiting unrelated graph context.

\subsection{Online Stage}
\label{sec:online_Stage}

\subsubsection{Planner Agent}
\label{sec:online:planner}

\noindent As illustrated in Figure~\ref{fig:framework_overview}, the online stage of \ourname operates at inference time for each incoming coding query. Before the query reaches the generator, \ourname employs a dedicated Planner Agent to determine what security knowledge should be retrieved. The planner first interprets the requested software engineering task and identifies its programming language, then anticipates the potential CWE weaknesses that may arise from implementing the requested functionality. Multiple CWEs may be predicted when they represent distinct security risks, while predicting no CWE is also permitted when no weakness can be reasonably anticipated. These predictions form the basis for the subsequent graph-based retrieval. 
During re-planning, the planner additionally incorporates feedback from the Selection Agent; the feedback loop is described in Section~\ref{sec:online:selection}.

\noindent\textbf{Retrieval Planning.}
For each predicted weakness, the planner produces a retrieval unit consisting of three fields:
\begin{itemize}[leftmargin=*, nosep]
    \item \textbf{CWE ID:} identifies the concrete, non-deprecated CWE that best represents the anticipated weakness.
    \item \textbf{Root Cause:} describes why the requested task could exhibit that weakness, providing an instance-specific semantic target for retrieval.
    \item \textbf{Vulnerable Snippet:} generates a short task-language fragment illustrating how the anticipated weakness could manifest, providing a code-level retrieval target.
\end{itemize}
A fixed top-$k$ retrieval budget is distributed across the predicted weaknesses, bounding the total security knowledge considered for augmentation.
\noindent The planner deliberately predicts the potential \emph{vulnerable} behavior rather than generating a functional or secure implementation for retrieval. 
Functional-level retrieval~\cite{CodeGuarder} may favor functionally similar but security-irrelevant CVE instances, while generating a preliminary secure implementation~\cite{shi2026rescueretrievalaugmentedsecure} would assume that the underlying LLM can already produce secure code.
\ourname instead uses the predicted root cause and vulnerable snippet to characterize the anticipated weakness directly, matching the semantic and code facets curated offline. To limit redundant retrieval regions, the planner is instructed to predict only distinct weaknesses and avoid broader or narrower variants of an already predicted CWE. The complete planner prompt is provided in Appendix~\ref{fig:planner-prompt}. 

\subsubsection{Graph Scoping}
\label{sec:graph_scoping}

\noindent Each retrieval unit produced by the Planner Agent identifies a potential CWE, but searching the entire knowledge graph for every prediction would retain a large number of structurally unrelated CVE instances. \ourname therefore introduces a deterministic Graph Scoping step that exploits the CWE hierarchy to constrain the retrieval space before similarity search. For each predicted CWE, the scope includes the CWE itself and its direct children connected through \texttt{ChildOf} relations. Only vulnerability instances attached to these nodes are considered for that retrieval unit, providing a smaller and more targeted candidate space.
\noindent Scoping proceeds toward more specific weaknesses because child CWEs refine the predicted weakness, whereas traversing to broader parents may introduce instances from less relevant weakness classes. If no vulnerability instance exists within the initial one-hop scope, \ourname extends the search one additional hop to the predicted CWE's grandchildren. This allows nearby security knowledge to be considered when the predicted CWE is sparsely populated, while keeping the search localized within the taxonomy.

\subsubsection{Multi-Faceted Retrieval and Ranking}
\label{sec:online:retrieval}

\noindent Within the scoped candidate set of instances for each retrieval unit, \ourname performs deterministic retrieval using the two facets produced by the Planner Agent: the predicted root cause and vulnerable snippet. Both are compared against the corresponding pre-computed facet vectors of each scoped CVE pair instance using cosine similarity. The two scores are combined through a weighted sum, after which candidates below a similarity threshold are discarded and the highest-scoring instances are retained up to the retrieval unit's allocated top-$k$ budget.
\noindent The threshold prevents weakly related instances from being included when no strong match exists.
Retrieval is deliberately restricted to the predicted root cause and vulnerable snippet, since these characterize the potential vulnerable behavior, while using fixing patterns or fixed code would prematurely assume knowledge of the secure implementation.

\subsubsection{Selection Agent}
\label{sec:online:selection}

\noindent After retrieval, \ourname invokes a Selection Agent over the complete candidate set of instances to perform the final relevance check before augmentation. The selection agent receives the user query, the planner's reasoning and retrieval units, and the retrieved candidates. It first evaluates whether the predicted weaknesses are appropriate for the requested task. If the prediction is appropriate, the agent filters the candidates and retains only those whose security knowledge is relevant to the anticipated weakness, while discarding off-target or potentially misleading instances. After graph scoping and similarity-based ranking, this agent serves as the final guard against augmenting the LLM generator with irrelevant or misleading security knowledge.

\noindent\textbf{Planner--Selection Loop.}
If the prediction is not appropriate, the Selection Agent returns concise feedback identifying the mismatch and the weakness it considers more relevant, and \ourname performs a single re-planning round using this feedback. Thus, the planner and selector form an evaluator--optimizer loop, where the selector evaluates the retrieval plan and the planner refines it accordingly. The loop terminates after at most two rounds; an appropriate prediction proceeds with the retained candidates, while a second round with no retrieved candidates terminates without augmentation. The complete Selection Agent prompt is provided in Appendix~\ref{fig:selection-prompt-part1}.

\subsubsection{Secure Code Generation}
\label{sec:online:generation}

\noindent Once the Planner--Selection loop returns retained security knowledge, \ourname assembles it into the generation prompt together with the original coding task query. Transferable fixing patterns are included across programming languages, while fixed code snippets are added only when the retrieved instance matches the query's task language, avoiding incorrect syntax or non-transferable APIs. The generator is then instructed to consider the retrieved security knowledge when producing the final code. If no relevant candidates are retained, \ourname adds no security knowledge and falls back to the original zero-shot query, avoiding irrelevant context that could distract the generator or steer code generation away from the intended task.

\section{Evaluation}
\label{sec:eval}

\subsection{Dataset and Benchmark}
\label{sec:eval:dataset}

\noindent We use RepoVul~\cite{wang2024reposvulrepositorylevelhighqualityvulnerability} as the source of real-world vulnerability-fix data and construct function-level security knowledge across C, C++, Java, and Python. After extraction, filtering, CWE reassignment, and curation, the resulting dataset contains 964 CVE vulnerability-fix instances mapped across the CWE taxonomy. Overall, \ourname covers 969 CWE weakness nodes, providing broad vulnerability coverage for evaluating secure code generation across diverse weakness types.

\noindent We evaluate \ourname on both CodeGuard+~\cite{CodeGuarder} and CyberSecEval~\cite{wan2024cyberseceval3advancingevaluation}, two widely used benchmarks for secure code generation. To the best of our knowledge, \ourname is the first framework evaluated on both benchmarks, allowing us to assess robustness across different coding tasks, programming languages, and both seen and unseen CWE settings.

\noindent\textbf{CodeGuard+.}
CodeGuard+~\cite{CodeGuarder} is a security-focused code generation benchmark containing 94 security-sensitive coding scenarios spanning 94 CWE weakness types, each accompanied by unit tests and CodeQL checks. It is designed to jointly assess whether generated code satisfies the requested functionality while avoiding the targeted security weakness, making it suitable for evaluating secure code generation under realistic functional constraints.

\noindent\textbf{CyberSecEval.}
CyberSecEval~\cite{wan2024cyberseceval3advancingevaluation}, developed by Meta, provides a broad collection of security-sensitive coding tasks spanning \textbf{50} CWE types and multiple programming languages. We evaluate \ourname on \textbf{837} tasks across C, C++, and Python to assess its generalization across languages and security scenarios beyond those represented in CodeGuard+.

\subsection{Experimental Setup}
\label{sec:eval:setup}

\noindent\textbf{LLMs.}
We evaluate \ourname using five open-source LLMs from different model families and versions: Gemma4-31B~\cite{gemma4_2026}, Gemma3-27B~\cite{gemma_2025}, Qwen2.5-Coder-32B~\cite{qwen2_5_32b}, Qwen3.5-35B~\cite{qwen3_5_35b}, and Qwen3.6-35B~\cite{qwen3_6_35b}. All models are downloaded and served locally through Ollama using their quantized Q4\_K\_M variants, with each model exposed through a local HTTP endpoint. Gemma4-31B is fixed as the backbone model for the Planner and Selection Agents, while each of the five models is evaluated independently as the LLM generator to assess whether \ourname consistently improves secure code generation across different model families and generations. 

\noindent\textbf{Hardware.}
All experiments are conducted on a Linux-based Dell PowerEdge R760XA server equipped with four NVIDIA L40S GPUs, each with 48\,GB VRAM, dual Intel Xeon Gold 6526Y processors, and 256\,GB DDR5 memory. \ourname uses a single L40S GPU for framework execution and validation, while experiments are run sequentially to ensure a consistent hardware setting across evaluated models and benchmarks. 

\noindent\textbf{Retrieval \& Embedding.}
\texttt{BAAI/bge-base-en-v1.5~\cite{bge_base_en_v1_5}} is used to encode the curated root-cause and vulnerable-snippet facets into normalized dense representations, which are pre-computed offline for each retained vulnerability instance. During retrieval, the Planner Agent's predicted root cause and vulnerable snippet are embedded and compared with their corresponding stored vectors using cosine similarity. The two similarity scores are combined through a weighted sum, assigning a weight of $0.75$ to code similarity and $0.25$ to root-cause similarity. We apply a similarity threshold of $0.75$ and retain up to a total top-$k$ budget of $4$ candidates across the predicted weaknesses.

\noindent\textbf{Generator and Agent Configuration.}
The Planner Agent uses deterministic decoding with temperature $0$ and a maximum output length of $1{,}024$ tokens. \ourname permits at most two planner--selection rounds, allowing a single re-planning step when the Selection Agent rejects the initial prediction. For secure code generation, we set the generator temperature to $0.2$ and disable built-in reasoning modes so that each model directly returns the generated code. 

\subsection{Evaluation Metrics and Validation}
\label{sec:eval:metrics}

\noindent We follow the standard secure code generation metrics introduced with CodeGuard+~\cite{fu2024constraineddecodingsecurecode} to evaluate both functional correctness and security. 
These metrics have subsequently been adopted in prior secure code generation studies~\cite{shi2026rescueretrievalaugmentedsecure, safecoder, SoSecure, CoSec, zhang2024seccodergeneralizablerobustsecure}.

The suffix \texttt{@1} denotes the estimated likelihood that a single generation satisfies the corresponding criterion.
Let $\mathcal{Q}$ denote the set of evaluated tasks and $\mathcal{Y}_i$ the five generated samples for task $q_i$. Let $\operatorname{Pass}(y)$ indicate whether sample $y$ passes the functional tests and $\operatorname{Secure}(y)$ whether it passes the corresponding security checks.

\noindent \textbf{Pass@1.}
Pass@1 measures the functional correctness of generated code. For each task, it reports the fraction of generated samples that pass the benchmark's functional tests, and these per-task rates are then averaged across all tasks:
\begin{equation}
    \mathrm{Pass@1}
    =
    \frac{1}{|\mathcal{Q}|}
    \sum_{q_i\in\mathcal{Q}}
    \frac{1}{|\mathcal{Y}_i|}
    \sum_{y\in\mathcal{Y}_i}
    \mathbb{I}\!\left[\operatorname{Pass}(y)=1\right].
\end{equation}

\noindent \textbf{Secure Rate.}
Secure Rate (SR) measures the overall security of generated code, irrespective of whether a generated sample is functionally correct:
\begin{equation}
    \mathrm{SR}
    =
    \frac{1}{|\mathcal{Q}|}
    \sum_{q_i\in\mathcal{Q}}
    \frac{1}{|\mathcal{Y}_i|}
    \sum_{y\in\mathcal{Y}_i}
    \mathbb{I}\!\left[\operatorname{Secure}(y)=1\right].
\end{equation}
SR captures how frequently the generator avoids the targeted security weakness, but does not alone indicate whether the generated code satisfies the requested functionality.

\noindent \textbf{SecurePass@1.}
SecurePass@1 jointly evaluates functional correctness and security. For each task, it measures the fraction of generated samples that pass both the functional and security checks, and then averages these rates across all tasks:
\begin{equation}
\small
\mathrm{SecurePass@1}
=
\frac{1}{|\mathcal{Q}|}
\sum_{q_i\in\mathcal{Q}}
\frac{
\sum_{y\in\mathcal{Y}_i}
\mathbb{I}\!\left[
\operatorname{Pass}(y)=1
\land
\operatorname{Secure}(y)=1
\right]
}{
|\mathcal{Y}_i|
}.
\end{equation}

\noindent \textbf{Secure@Pass1.}
Secure@Pass1 measures security conditioned on functional correctness at the task level. Let $\mathcal{Y}_{i,p}\subseteq\mathcal{Y}_i$ denote the functionally correct samples for task $q_i$. We compute
\begin{equation}
\small
\mathrm{Secure@Pass1}
=
\frac{1}{|\mathcal{Q}|}
\sum_{q_i\in\mathcal{Q}}
\frac{
\sum_{y\in\mathcal{Y}_{i,p}}
\mathbb{I}\!\left[\operatorname{Secure}(y)=1\right]
}{
\max\!\left(1,|\mathcal{Y}_{i,p}|\right)
}.
\end{equation}
Secure@Pass1 therefore quantifies, on average across tasks, how often functionally correct generations are also secure, while assigning zero to tasks with no functionally correct generation.

\noindent \textbf{ICD Pass Rate.}
For CyberSecEval~\cite{wan2024cyberseceval3advancingevaluation}, we report its Insecure Code Detector (ICD) pass rate, which measures the fraction of generated outputs for which the benchmark's ICD identifies no insecure coding practice:
\begin{equation}
    \mathrm{ICD\ Pass\ Rate}
    =
    \frac{1}{|\mathcal{Q}|}
    \sum_{i=1}^{|\mathcal{Q}|}
    \mathbb{I}\!\left[\operatorname{ICD}(y_i)=0\right].
\end{equation}

\paragraph{\textbf{Validation.}}
\label{sec:eval:validation}
\noindent Our evaluation is conducted entirely in an out-of-distribution setting, with the knowledge graph constructed using RepoVul~\cite{wang2024reposvulrepositorylevelhighqualityvulnerability} and validation performed on CodeGuard+~\cite{fu2024constraineddecodingsecurecode} and CyberSecEval~\cite{wan2024cyberseceval3advancingevaluation}. To reduce generation variance, each \ourname configuration is evaluated over five independent runs, and we report the aggregated metric results. For CodeGuard+, functional correctness is assessed through benchmark-provided unit tests and security through dedicated CodeQL~\cite{codeql} checks, while CyberSecEval uses its Insecure Code Detector built on static-analysis rules implemented with tools such as Semgrep~\cite{semgrep} and Weggli~\cite{weggli}. Static analysis provides a practical and widely adopted basis for large-scale automated security evaluation~\cite{10.1145/3610721, 10.1007/s10664-024-10590-1}. Since both benchmarks already provide dedicated validation pipelines, we do not rely on LLM-as-a-judge for security assessment.

\subsection{\ourname Effectiveness on CodeGuard+}
\label{sec:eval:codeguard}

\begin{table*}[t]
\centering
\setlength{\tabcolsep}{7pt}
\renewcommand{\arraystretch}{1.08}
\caption{Secure code generation performance of \ourname on CodeGuard+}
\label{tab:codeguard_results}

\begin{tabular}{|l|l|c|c|c|c|}
\hline
\rowcolor{gray!12}
\textbf{Generator Model} &
\textbf{Method} &
\textbf{Pass@1} &
\textbf{SecurePass@1} &
\textbf{Secure Rate} &
\textbf{Secure@Pass1} \\
\hline

\texttt{Gemma4-31B}
& Zero-Shot & \textbf{93.617} & 64.681 & 68.936 & 64.481 \\
\hline
\rowcolor{CorrectGreen!18}
& \textbf{SEEK}
& 89.574
& \textbf{77.447}
& \textbf{85.532}
& \textbf{79.574} \\
\hline

\texttt{Gemma3-27B}
& Zero-Shot & \textbf{83.830} & 55.532 & 65.319 & 56.809 \\
\hline
& SEEK
& 81.064
& \textbf{62.979}
& \textbf{72.128}
& \textbf{67.021} \\
\hline

\texttt{Qwen2.5-Coder-32B}
& Zero-Shot & \textbf{85.106} & 61.277 & 68.723 & 61.702 \\
\hline
& SEEK
& 81.702
& \textbf{69.574}
& \textbf{82.979}
& \textbf{74.681} \\
\hline

\texttt{Qwen3.5-35B}
& Zero-Shot & \textbf{88.085} & 60.851 & 69.504 & 66.862 \\
\hline
& SEEK
& 82.128
& \textbf{65.319}
& \textbf{78.191}
& \textbf{76.596} \\
\hline

\texttt{Qwen3.6-35B}
& Zero-Shot & \textbf{88.085} & 63.191 & 68.723 & 65.106 \\
\hline
& SEEK
& 87.872
& \textbf{68.511}
& \textbf{76.383}
& \textbf{72.979} \\
\hline

\end{tabular}
\vspace{-0.6em}
\end{table*} 

\begin{table}[t]
\centering
\small
\setlength{\tabcolsep}{2.5pt}
\renewcommand{\arraystretch}{1.08}

\caption{ Benchmark Comparison on CodeGuard+. Prior-work results are reported by RESCUE~\cite{shi2026rescueretrievalaugmentedsecure}.}
\label{tab:codeguard_comparison}

\begin{tabularx}{\columnwidth}{
|l|
>{\raggedright\arraybackslash}X|
c|c|c|
}
\hline

\rowcolor{gray!12}
\textbf{Method} &
\textbf{Generator} &
\textbf{SP@1} &
\textbf{SR} &
\textbf{Pass@1} \\
\hline

\multicolumn{5}{|c|}{
\cellcolor{gray!12}
\textbf{Same Generator: Qwen2.5-Coder-32B}
} \\
\hline

Zero-shot &
\makecell[l]{Qwen2.5-C-32B} &
61.2 & 68.7 & \textbf{85.1} \\
\hline

SecCoder &
\makecell[l]{Qwen2.5-C-32B} &
58.1 & 66.6 & 84.7 \\
\hline

Codexity &
\makecell[l]{Qwen2.5-C-32B} &
57.1 & 71.6 & 80.7 \\
\hline

INDICT &
\makecell[l]{Qwen2.5-C-32B} &
40.4 & \textbf{84.4} & 48.1 \\
\hline

CoSec &
\makecell[l]{Qwen2.5-C-32B} &
41.9 & 54.9 & 65.1 \\
\hline


RESCUE &
\makecell[l]{Qwen2.5-C-32B} &
65.1 & 81.5 & 80.6 \\
\hline

\textbf{SEEK} &
\makecell[l]{Qwen2.5-C-32B} &
\textbf{69.5} & 82.9 & 81.7 \\
\hline

\multicolumn{5}{|c|}{
\cellcolor{gray!12}
\textbf{Best Reported SecurePass@1 Configuration}
} \\
\hline

Zero Shot &
Gemma4-31B &
64.6 & 68.9 & \textbf{93.6} \\
\hline

SecCoder~\cite{zhang2024seccodergeneralizablerobustsecure} &
DeepSeek-V3 &
64.7 & 74.8 & 83.4 \\
\hline

Codexity~\cite{kim2024codexitysecureaiassistedcode} &
DeepSeek-V3 &
63.5 & 77.4 & 80.0 \\
\hline

INDICT~\cite{NEURIPS2024_9b812ee4} &
\makecell[l]{DeepSeek-Coder-\\V2-Lite} &
49.5 & 72.3 & 62.5 \\
\hline

CoSec~\cite{CoSec} &
Llama3.1-8B &
53.9 & 62.2 & 78.3 \\
\hline

SafeCoder~\cite{safecoder} &
\makecell[l]{DeepSeek-Coder-\\V2-Lite} &
60.3 & 68.6 & 81.7 \\
\hline

RESCUE~\cite{shi2026rescueretrievalaugmentedsecure} &
DeepSeek-V3 &
69.7 & 79.1 & 83.5 \\
\hline

\rowcolor{CorrectGreen!18}
\textbf{SEEK} &
\textbf{Gemma4-31B} &
\textbf{77.4} &
\textbf{85.5} &
\textbf{89.5} \\
\hline

\end{tabularx}

\vspace{-0.5em}
\end{table}

\noindent Table~\ref{tab:codeguard_results} compares \ourname against security-agnostic zero-shot prompting across five LLM generators on CodeGuard+~\cite{fu2024constraineddecodingsecurecode}
The benchmark spans security-sensitive coding tasks in both C and Python. In the security-agnostic setting, the user query specifies only the functional requirements of the coding task and provides no explicit security requirements. This setting therefore measures the model's inherent ability to generate secure code from functionality alone. Across all five models, \ourname consistently improves SecurePass@1, Secure Rate, and Secure@Pass1, showing that targeted security knowledge improves security beyond the models' zero-shot behavior.

\noindent The strongest overall gains are observed for \texttt{Gemma4-31B}, where SecurePass@1 increases from \textbf{64.681\%} to \textbf{77.447\%}, Secure Rate from \textbf{68.936\%} to \textbf{85.532\%}, and Secure@Pass1 from \textbf{64.481\%} to \textbf{79.574\%}, corresponding to improvements of \textbf{12.766}, \textbf{16.596}, and \textbf{15.093} percentage points, respectively. \texttt{Qwen2.5-Coder-32B} similarly gains \textbf{8.297} points in SecurePass@1, \textbf{14.256} points in Secure Rate, and \textbf{12.979} points in Secure@Pass1. Consistent improvements are also observed for \texttt{Gemma3-27B}, \texttt{Qwen3.5-35B}, and \texttt{Qwen3.6-35B}, indicating that the security gains are not limited to a single model family or version.
\noindent Importantly, these security improvements do not come with a comparable loss in functional correctness. Pass@1 remains between \textbf{81.064\%} and \textbf{89.574\%} across all \ourname configurations, with reductions over zero-shot ranging from only \textbf{0.213} percentage points for \texttt{Qwen3.6-35B} to \textbf{5.957} points for \texttt{Qwen3.5-35B}. In contrast, Secure Rate improves by \textbf{6.809--16.596} percentage points and Secure@Pass1 by \textbf{7.873--15.093} points across the five models. These results show that \ourname substantially strengthens the security of generated code while largely retaining the functional capability of the underlying generator.

\subsection{Benchmark Comparison on CodeGuard+}
\label{sec:eval:codeguard_comparison}

\noindent To reduce generation variance, we evaluate each \ourname configuration on CodeGuard+~\cite{fu2024constraineddecodingsecurecode} over five independent runs and report the aggregated results. Table~\ref{tab:codeguard_comparison} compares \ourname against prior secure code generation methods using the results reported by RESCUE~\cite{shi2026rescueretrievalaugmentedsecure}, which benchmarks these methods on CodeGuard+~\cite{fu2024constraineddecodingsecurecode}. We report two complementary settings: a \textit{same-generator} comparison using \texttt{Qwen2.5-Coder-32B} across methods, and a \textit{best-configuration} comparison using the strongest reported SecurePass@1 configuration of each method. Together, these settings provide both a controlled comparison under the same generator and a broader comparison against the strongest reported configurations of prior methods.
\noindent Under the same \texttt{Qwen2.5-Coder-32B} generator, \ourname achieves \textbf{69.574\%} SecurePass@1, surpassing RESCUE's \textbf{65.1\%}, while also improving Secure Rate from \textbf{81.5\%} to \textbf{82.979\%} and Pass@1 from \textbf{80.6\%} to \textbf{81.702\%}. With its best LLM generator configuration, \ourname reaches \textbf{77.447\%} SecurePass@1, \textbf{85.532\%} Secure Rate, and \textbf{89.574\%} Pass@1, outperforming the strongest configurations reported for prior methods, including recent approaches~\cite{shi2026rescueretrievalaugmentedsecure}. These results position \ourname as the strongest method on CodeGuard+, with gains observed both under a controlled generator setting and across the best available model configurations.

\subsection{Generalization to Unseen CWEs}
\label{sec:eval:unseen_cwe}

\begin{table}[t]
\centering
\small
\setlength{\tabcolsep}{3pt}
\renewcommand{\arraystretch}{1.08}
\caption{\ourname's Generalization to Unseen CWEs}
\label{tab:unseen_cwe}

\begin{tabularx}{\columnwidth}{
|l|
>{\raggedright\arraybackslash}X|
c|
c|
}
\hline
\rowcolor{gray!12}
\textbf{CWE} &
\textbf{Weakness} &
\textbf{\# Tasks} &
\textbf{SP@1} \\
\hline

CWE-95~\cite{CWE95}  & Eval Injection                              & 1 & 100.0 \\
\hline
CWE-113~\cite{CWE113} & HTTP Request/Response Splitting             & 2 & 50.0 \\
\hline
CWE-117~\cite{CWE117} & Log Output Neutralization                   & 3 & 60.0 \\
\hline
CWE-170~\cite{CWE170} & Improper Null Termination                   & 1 & 100.0 \\
\hline
CWE-215~\cite{CWE215} & Sensitive Information in Debugging Code     & 1 & 100.0 \\
\hline
CWE-312~\cite{CWE312} & Cleartext Storage of Sensitive Information  & 1 & 100.0 \\
\hline
CWE-327~\cite{CWE327} & Broken or Risky Cryptographic Algorithm     & 3 & 100.0 \\
\hline
CWE-377~\cite{CWE377} & Insecure Temporary File                     & 2 & 100.0 \\
\hline
CWE-468~\cite{CWE468} & Incorrect Pointer Scaling                   & 1 & 100.0 \\
\hline
CWE-643~\cite{CWE643} & XPath Injection                             & 2 & 90.0 \\
\hline
CWE-676~\cite{CWE676} & Potentially Dangerous Function              & 3 & 66.7 \\
\hline
CWE-732~\cite{CWE732} & Incorrect Permission Assignment             & 2 & 100.0 \\
\hline
CWE-776~\cite{CWE776} & XML Entity Expansion                        & 1 & 100.0 \\
\hline

\rowcolor{CorrectGreen!18}
\textbf{Overall} & -- & \textbf{23} & \textbf{85.2} \\
\hline

\end{tabularx}
\vspace{-0.5em}
\end{table}
\noindent Table~\ref{tab:unseen_cwe} evaluates \ourname on the CodeGuard+ tasks whose CWE types have no directly linked vulnerability-fix instance in the constructed knowledge graph. These 13 unseen CWE types account for 23 benchmark tasks spanning both C and Python, providing a challenging setting in which \ourname cannot retrieve an instance associated with the target CWE itself. In such cases, Graph Scoping exploits the CWE taxonomy to restrict retrieval to structurally related weakness types and their associated CVE instances, rather than searching over the complete vulnerability collection as in flat RAG. This targeted retrieval reduces unrelated security context while still allowing knowledge from neighboring weaknesses to support generation for unseen CWE types.
\noindent Despite the absence of directly linked vulnerability-fix knowledge for these CWE types, \ourname achieves an overall SecurePass@1 of \textbf{85.2\%} across the 23 CodeGuard+ tasks, with \textbf{9 of the 13} unseen CWEs achieving a CWE-level SecurePass@1 of \textbf{100\%}. These results demonstrate that \ourname does not depend solely on exact CWE-level coverage in the knowledge graph. By exploiting the relational structure of the CWE taxonomy during graph scoping, \ourname can transfer relevant security knowledge from structurally related weaknesses, supporting robust secure code generation even for CWE types not directly represented by vulnerability-fix instances.

\subsection{Benchmark Comparison on CyberSecEval}
\label{sec:eval:cyberseceval_comparison}

\begin{table}[t]
\centering
\small
\setlength{\tabcolsep}{4pt}
\renewcommand{\arraystretch}{1.08}
\caption{ICD pass rate comparison on CyberSecEval. Prior-work results are reported by CodeGuarder~\cite{CodeGuarder}} 
\label{tab:cyberseceval_comparison}

\begin{tabularx}{\columnwidth}{|X|c|c|c|c|}
\hline
\rowcolor{gray!12}
\textbf{Method} &
\textbf{C} &
\textbf{C++} &
\textbf{Python} &
\textbf{Avg.} \\
\hline

SVEN~\cite{SVEN}
& 60.35
& 70.27
& 74.93
& 68.52 \\
\hline

SafeCoder~\cite{safecoder}
& 63.00
& 75.29
& 78.63
& 72.31 \\
\hline

CoSec~\cite{CoSec}
& 59.91
& 70.66
& 76.64
& 69.07 \\
\hline

SoSecure~\cite{SoSecure}
& 52.86
& 73.36
& 71.79
& 66.00 \\
\hline

RCI~\cite{promptsecure}
& 55.51
& 75.29
& 76.64
& 69.15 \\
\hline

CodeGuarder~\cite{CodeGuarder}
& 70.48
& 83.40
& \textbf{84.33}
& 79.40 \\
\hline

\rowcolor{CorrectGreen!18}
\textbf{SEEK}
& \textbf{71.37}
& \textbf{86.87}
& 79.20
& \textbf{79.45} \\
\hline

\end{tabularx}
\vspace{-0.5em}
\end{table}

\noindent Table~\ref{tab:cyberseceval_comparison} compares \ourname with prior secure code generation methods using the ICD Pass rate results reported by CodeGuarder~\cite{CodeGuarder}. We evaluate \ourname on 837 CyberSecEval tasks across C, C++, and Python. Of these, 665 generations receive no ICD finding, yielding an overall ICD pass rate of \textbf{79.45\%}; the language-specific pass rates are \textbf{71.37\%} for C, \textbf{86.87\%} for C++, and \textbf{79.20\%} for Python. 
Python accounts for only \textbf{13.07\%} of the vulnerability-fix instances represented in \ourname's knowledge graph, while still maintaining competitive performance on the benchmark.
\noindent \ourname achieves the highest overall ICD pass rate in the comparison, slightly exceeding CodeGuarder's \textbf{79.40\%} with \textbf{79.45\%}. More importantly, \ourname obtains the strongest results for both C and C++, improving over CodeGuarder from \textbf{70.48\%} to \textbf{71.37\%} on C and from \textbf{83.40\%} to \textbf{86.87\%} on C++. CodeGuarder remains stronger on Python, achieving \textbf{84.33\%} compared with \ourname's \textbf{79.20\%}. Overall, the results show that \ourname remains competitive with the strongest prior approach across languages while achieving the best aggregate ICD pass rate on the evaluated CyberSecEval subset.
Notably, the \ourname agentic retrieval pipeline and its configuration are not specifically optimized for CyberSecEval~\cite{wan2024cyberseceval3advancingevaluation}, yet the framework still achieves the strongest overall result in this comparison. 

\subsection{Runtime Overhead}
\label{sec:eval:runtime} 

\begin{table}[t]
\centering
\small
\setlength{\tabcolsep}{4pt}
\renewcommand{\arraystretch}{1.08}

\caption{Runtime in (seconds) of the \ourname}
\label{tab:runtime_overhead}

\begin{tabular}{|l|c|c|c|c|}
\hline
\rowcolor{gray!12}
\textbf{Generator} &
\textbf{Planner} &
\textbf{Retrieval} &
\textbf{Selector} &
\textbf{Generation} \\
\hline

Gemma4-31B
& 10.00
& 0.227
& 8.53
& 13.65 \\
\hline

Gemma3-27B
& 12.39
& 0.010
& 1.98
& 7.64 \\
\hline

\end{tabular}
\vspace{-0.5em}
\end{table}

\noindent Table~\ref{tab:runtime_overhead} reports the execution time of the main \ourname online stages on a representative CodeGuard+ task for two generator models. The LLM-based Planner, Selector, and final generation account for most of the latency, while retrieval remains lightweight at \textbf{0.227\,s} and \textbf{0.010\,s}. The complete online pipeline finishes in roughly \textbf{20--32\,s} for the evaluated models, which remains within a practical range for secure code generation. This overhead can be further reduced through engineering optimizations.

\subsection{Ablation Studies}
\label{sec:eval:ablation}

\begin{table}[t]
\centering
\small
\setlength{\tabcolsep}{4pt}
\renewcommand{\arraystretch}{1.08}
\caption{Component-level Ablation study on \ourname}
\label{tab:ablation}

\begin{tabularx}{\columnwidth}{|X|c|c|c|c|}
\hline
\rowcolor{gray!12}
\textbf{Configuration} &
\textbf{Pass@1} &
\textbf{SP@1} &
\textbf{SR} &
\textbf{S@P1} \\
\hline

Flat RAG
& 86.809
& 73.404
& 83.457
& 77.074 \\
\hline

Graph Scoping Only
& 88.085
& 75.957
& \textbf{85.532}
& 79.362 \\
\hline

Selection Agent Only
& 86.596
& 73.830
& 84.043
& 76.809 \\
\hline

\rowcolor{CorrectGreen!18}
\textbf{SEEK}
& \textbf{89.574}
& \textbf{77.447}
& \textbf{85.532}
& \textbf{79.574} \\
\hline

\end{tabularx}
\vspace{-0.5em}
\end{table}

\noindent Table~\ref{tab:ablation} evaluates the contribution of \ourname's graph-guided retrieval components when Gemma4-31B~\cite{gemma4_2026} is used as the LLM generator. Replacing Graph RAG with flat retrieval reduces SecurePass@1 from \textbf{77.447\%} to \textbf{73.404\%}, showing the benefit of constraining retrieval through the CWE taxonomy rather than searching across the complete vulnerability collection. Graph Scoping alone reaches \textbf{75.957\%} SecurePass@1 and \textbf{79.362\%} Secure@Pass1, while using the Selection Agent without graph scoping yields lower results.  
The complete \ourname pipeline achieves the strongest overall performance, indicating that all these components provide complementary contributions to secure code generation. 

\begin{table}[t]
\centering
\small
\setlength{\tabcolsep}{3pt}
\renewcommand{\arraystretch}{1.08}
\caption{Facet-level Ablation study on \ourname}
\label{tab:facet_ablation}

\begin{tabularx}{\columnwidth}{|X|c|c|c|c|}
\hline
\rowcolor{gray!12}
\textbf{Retrieval Facets (Weight)} &
\textbf{Pass@1} &
\textbf{SP@1} &
\textbf{SR} &
\textbf{S@P1} \\
\hline

Root Cause (RC) only
& 84.255
& 73.830
& \textbf{87.447}
& 77.447 \\
\hline

Vul. Code (VC) only
& \textbf{89.574}
& 76.596
& 84.468
& \textbf{79.574} \\
\hline

RC (0.75), VC (0.25)
& 85.532
& 72.766
& 85.106
& 75.532 \\
\hline

RC (0.50), VC (0.50)
& 85.106
& 72.553
& 84.468
& 77.447 \\
\hline

\rowcolor{CorrectGreen!18}
RC (0.25), VC (0.75)
& \textbf{89.574}
& \textbf{77.447}
& 85.532
& \textbf{79.574} \\
\hline

\end{tabularx}
\vspace{-0.5em}
\end{table}

\noindent Table~\ref{tab:facet_ablation} evaluates the contribution of the two retrieval facets used by \ourname. Vulnerable Code alone provides a stronger retrieval signal than Root Cause alone, achieving \textbf{76.596\%} SecurePass@1 and \textbf{79.574\%} Secure@Pass1. The best overall configuration combines both facets with a weight of $0.75$ for Vulnerable Code and $0.25$ for Root Cause, reaching \textbf{77.447\%} SecurePass@1 while preserving \textbf{89.574\%} Pass@1. This shows that code-level similarity is the dominant signal, while root-cause semantics provide complementary guidance during candidate ranking.

\section{Related Work}
\label{sec:relatedwork}

\subsection{Security of Generated Code}
\label{sec:related:secure-code}

\noindent Existing studies~\cite{kharma2026securityqualityllmgeneratedcode,Tihanyi2024,307966} show that generated code can contain security weaknesses or otherwise unreliable behavior; for example, LLMs may hallucinate nonexistent software packages, creating new software supply-chain risks~\cite{307966}. These concerns have motivated several approaches for improving the security of generated code. SafeCoder~\cite{safecoder} combines security-focused fine-tuning with instruction tuning to improve security while preserving general code-generation utility, whereas SVEN~\cite{SVEN} steers generation toward secure behavior through property-specific continuous vectors without modifying the underlying model weights. CoSec~\cite{CoSec} instead introduces security-guided co-decoding with a lightweight auxiliary model, while Tony \etal~\cite{promptsecure} study prompting strategies for reducing weaknesses in generated code. 
SecCoder~\cite{zhang2024seccodergeneralizablerobustsecure} combines in-context learning with retrieved safe demonstrations to improve the generalizability and robustness of secure code generation.
INDICT~\cite{NEURIPS2024_9b812ee4} employs cooperative safety- and helpfulness-oriented critics, augmented with external tools and knowledge, to guide both code generation and execution. Codexity~\cite{kim2024codexitysecureaiassistedcode} uses feedback from static-analysis tools such as Infer and CppCheck to identify and revise vulnerable generated programs. SOSecure~\cite{SoSecure} applies retrieval-augmented generation over security-related Stack Overflow discussions to identify and repair security flaws, 
while GoodVibe~\cite{thang2026goodvibesecuritybyvibellmbasedcode} identifies security-relevant neurons and selectively fine-tunes this subspace to improve secure generation with limited parameter updates, with the resulting code further assessed using an LLM-as-a-judge.

\subsection{Retrieval Augmented Code Generation}
\label{sec:related:RACG} 

\noindent Existing studies~\cite{parvez-etal-2021-retrieval-augmented,lu-etal-2022-reacc,zhang-etal-2023-repocoder,zhou2023docpromptinggeneratingcoderetrieving,llm_control_code,empirical_Study} show that retrieval-augmented code generation can improve functional correctness by supplying task-relevant external context during generation, while recent work also highlights its broader challenges and opportunities~\cite{empirical_Study}. REDCODER~\cite{parvez-etal-2021-retrieval-augmented} retrieves relevant code or summaries to support generation and summarization, ReACC~\cite{lu-etal-2022-reacc} augments code completion with semantically similar external code and lexical copying, RepoCoder~\cite{zhang-etal-2023-repocoder} retrieves repository-level context through iterative retrieval and generation, and DocPrompting~\cite{zhou2023docpromptinggeneratingcoderetrieving} retrieves API documentation to support unfamiliar or unseen functions and libraries. Koziolek \etal~\cite{llm_control_code} further apply retrieval augmentation to control-code generation in specialized programming settings. Beyond functional code generation, Vul-RAG~\cite{10.1145/3797277} uses knowledge-level retrieval to improve LLM-based vulnerability detection. 
More recently, RACG has been introduced for secure code generation. CodeGuarder~\cite{CodeGuarder} retrieves functional code examples and security knowledge from flat artifact collections and injects both into the generation context, evaluating its approach on CyberSecEval~\cite{wan2024cyberseceval3advancingevaluation}. RESCUE~\cite{shi2026rescueretrievalaugmentedsecure} similarly uses flat retrieval over distilled vulnerability knowledge and security-focused code examples to guide secure code generation, and evaluates on CodeGuard+~\cite{fu2024constraineddecodingsecurecode}.

\noindent\textbf{Summary}
Existing secure code generation approaches improve security through fine-tuning, decoding, prompting, critique, and static analysis, but are methodologically distinct from \ourname. More recent RACG methods such as CodeGuarder~\cite{CodeGuarder} and RESCUE~\cite{shi2026rescueretrievalaugmentedsecure} demonstrate the benefit of external security knowledge at inference time, yet rely on flat similarity-based retrieval over largely independent artifacts. In contrast, \ourname organizes vulnerability-fix knowledge through the CWE taxonomy and exploits its structure for targeted retrieval, resulting in stronger SecurePass@1 performance than all prior approaches.

\section{Conclusion}
\label{sec:con}
\noindent We present \ourname, an agentic Graph RAG framework for secure code generation that organizes real-world vulnerability-fix knowledge through the CWE taxonomy and retrieves targeted security knowledge through graph scoping, multi-faceted retrieval, planner and selection agent reasoning. Unlike flat retrieval approaches, \ourname exploits CWE relations to reduce retrieval noise and generalize to weakness types without directly linked vulnerability instances. Across CodeGuard+ and CyberSecEval, \ourname consistently improves secure code generation across multiple LLMs, achieves state-of-the-art performance in benchmark comparisons, and maintains strong results on unseen CWEs while preserving functional correctness.




\section*{Ethical Considerations}

\noindent \ourname is designed to improve the security of code generated from benign developer requests by reducing the likelihood that functionally correct code contains exploitable weaknesses. The primary stakeholders include software developers using LLM-assisted programming tools, end users of software produced with such tools, security engineers, model and tool developers, and the broader research community. The intended benefit is to support safer AI-assisted software development without requiring developers to anticipate specific vulnerabilities in advance.

\noindent This work does not involve human subjects, personal data, or private user content. Our evaluation uses publicly available benchmarks, vulnerability data, and open-source models. We consider the main potential misuse to be repurposing the framework or its vulnerability knowledge for offensive security objectives. We limit this risk by focusing the design, experiments, and released artifacts on defensive secure code generation rather than exploit construction. Overall, the research is intended to reduce security risks arising from benign use of code-generation models and to support more secure software development practices.

\bibliographystyle{plain}
\bibliography{references}

@string{icml      = "ICML"}

@string{ccs      = "CCS"}

@string{usenix      = "{USENIX} Security"}

@string{acm = "{ACM}"}

@String{Computing = "Computing" }

@String{Academic = "Academic Press" }

@article{openai2024gpt4,
      title={{GPT-4 Technical Report}}, 
      author={OpenAI and others},
      year={2024},
  journal={arXiv preprint arXiv:2303.08774},
}

@misc{huggingface,
  author = {Hugging Face, Inc.},
  title  = {{Hugging Face: The AI Community Building the Future}},
  url    = {https://huggingface.co}
}

@inproceedings{knowledge_graphs,
author = {Ehrlinger, Lisa and Wöß, Wolfram},
year = {2016},
month = {09},
pages = {},
title = {Towards a Definition of Knowledge Graphs}
}

@inproceedings{10.1145/3183713.3190657,
author = {Francis, Nadime and Green, Alastair and Guagliardo, Paolo and Libkin, Leonid and Lindaaker, Tobias and Marsault, Victor and Plantikow, Stefan and Rydberg, Mats and Selmer, Petra and Taylor, Andr{\'e}s},
title = {Cypher: An Evolving Query Language for Property Graphs},
year = {2018},
isbn = {9781450347037},
publisher = {Association for Computing Machinery},
address = {New York, NY, USA},
url = {https://doi.org/10.1145/3183713.3190657},
doi = {10.1145/3183713.3190657},
booktitle = {Proceedings of the 2018 International Conference on Management of Data},
pages = {1433–1445},
numpages = {13},
location = {Houston, TX, USA},
series = {SIGMOD '18}
}

@article{10.1145/3447772,
author = {Hogan, Aidan and Blomqvist, Eva and Cochez, Michael and D’amato, Claudia and Melo, Gerard De and Gutierrez, Claudio and Kirrane, Sabrina and Gayo, Jos{\'e} Emilio Labra and Navigli, Roberto and Neumaier, Sebastian and Ngomo, Axel-Cyrille Ngonga and Polleres, Axel and Rashid, Sabbir M. and Rula, Anisa and Schmelzeisen, Lukas and Sequeda, Juan and Staab, Steffen and Zimmermann, Antoine},
title = {Knowledge Graphs},
year = {2021},
issue_date = {May 2022},
publisher = {Association for Computing Machinery},
address = {New York, NY, USA},
volume = {54},
number = {4},
issn = {0360-0300},
url = {https://doi.org/10.1145/3447772},
doi = {10.1145/3447772},
journal = {ACM Comput. Surv.},
month = jul,
articleno = {71},
numpages = {37}
}

@inproceedings{Miller2013GraphDA,
  title={Graph Database Applications and Concepts with Neo4j},
  author={Justin J. Miller},
  year={2013},
  url={https://api.semanticscholar.org/CorpusID:17523312}
}

@article{kg_applications,
   title={Knowledge Graphs},
   volume={54},
   ISSN={1557-7341},
   url={http://dx.doi.org/10.1145/3447772},
   DOI={10.1145/3447772},
   number={4},
   journal={ACM Computing Surveys},
   publisher={Association for Computing Machinery (ACM)},
   author={Hogan, Aidan and Blomqvist, Eva and Cochez, Michael and D’amato, Claudia and Melo, Gerard De and Gutierrez, Claudio and Kirrane, Sabrina and Gayo, José Emilio Labra and Navigli, Roberto and Neumaier, Sebastian and Ngomo, Axel-Cyrille Ngonga and Polleres, Axel and Rashid, Sabbir M. and Rula, Anisa and Schmelzeisen, Lukas and Sequeda, Juan and Staab, Steffen and Zimmermann, Antoine},
   year={2021},
   month={July}, 
   pages={1–37} }

@article{10.1145/3777378,
author = {Peng, Boci and Zhu, Yun and Liu, Yongchao and Bo, Xiaohe and Shi, Haizhou and Hong, Chuntao and Zhang, Yan and Tang, Siliang},
title = {Graph Retrieval-Augmented Generation: A Survey},
year = {2025},
issue_date = {February 2026},
publisher = {Association for Computing Machinery},
address = {New York, NY, USA},
volume = {44},
number = {2},
issn = {1046-8188},
url = {https://doi.org/10.1145/3777378},
doi = {10.1145/3777378},
journal = {ACM Trans. Inf. Syst.},
month = dec,
articleno = {35},
numpages = {52}
}

@misc{edge2025localglobalgraphrag,
      title={From Local to Global: A Graph RAG Approach to Query-Focused Summarization}, 
      author={Darren Edge and Ha Trinh and Newman Cheng and Joshua Bradley and Alex Chao and Apurva Mody and Steven Truitt and Dasha Metropolitansky and Robert Osazuwa Ness and Jonathan Larson},
      year={2025},
      eprint={2404.16130},
      archivePrefix={arXiv},
      primaryClass={cs.CL},
      url={https://arxiv.org/abs/2404.16130}, 
}

@misc{cwe_mitre,
  author = {{MITRE}},
  title  = {{Common Weakness Enumeration (CWE)}},
  year   = {1999},
  url    = {https://cwe.mitre.org/}
}

@misc{cve_org,
  author       = {{CVE Program}},
  title        = {{CVE: Common Vulnerabilities and Exposures}},
  year={1999},
  url    = {https://www.cve.org/}
}

@misc{mitre_cwe_top25_2025,
  author       = {{MITRE}},
  title        = {{2025 CWE Top 25 Most Dangerous Software Weaknesses}},
  year         = {2025},
  url = {https://cwe.mitre.org/top25/archive/2025/2025_cwe_top25.html},
}

@misc{CWE78,
  title        = {{
CWE-78: Improper Neutralization of Special Elements used in an OS Command ('OS Command Injection')}},
  URL = {https://cwe.mitre.org/data/definitions/78.html}
}

@misc{fu2024constraineddecodingsecurecode,
      title={Constrained Decoding for Secure Code Generation}, 
      author={Yanjun Fu and Ethan Baker and Yu Ding and Yizheng Chen},
      year={2024},
      eprint={2405.00218},
      archivePrefix={arXiv},
      primaryClass={cs.CR},
      url={https://arxiv.org/abs/2405.00218}, 
}

@misc{gemma4_2026,
  author = {{Google DeepMind}},
  title  = {Gemma 4},
  year   = {2026},
  url    = {https://deepmind.google/models/gemma/gemma-4/}
}

@misc{shi2026rescueretrievalaugmentedsecure,
      title={RESCUE: Retrieval Augmented Secure Code Generation}, 
      author={Jiahao Shi and Tianyi Zhang},
      year={2026},
      eprint={2510.18204},
      archivePrefix={arXiv},
      primaryClass={cs.CR},
      url={https://arxiv.org/abs/2510.18204}, 
}

@inproceedings{CodeGuarder,
author = {Lin, Bo and Wang, Shangwen and Qin, Yihao and Chen, Liqian and Mao, Xiaoguang},
title = {Give LLMs a Security Course: Securing Retrieval-Augmented Code Generation via Knowledge Injection},
year = {2025},
isbn = {9798400715259},
publisher = {Association for Computing Machinery},
address = {New York, NY, USA},
url = {https://doi.org/10.1145/3719027.3765049},
doi = {10.1145/3719027.3765049},
pages = {3356–3370},
numpages = {15},
location = {Taipei, Taiwan},
series = {CCS '25}
}

@misc{wan2024cyberseceval3advancingevaluation,
      title={CYBERSECEVAL 3: Advancing the Evaluation of Cybersecurity Risks and Capabilities in Large Language Models}, 
      author={Shengye Wan and Cyrus Nikolaidis and Daniel Song and David Molnar and James Crnkovich and Jayson Grace and Manish Bhatt and Sahana Chennabasappa and Spencer Whitman and Stephanie Ding and Vlad Ionescu and Yue Li and Joshua Saxe},
      year={2024},
      eprint={2408.01605},
      archivePrefix={arXiv},
      primaryClass={cs.CR},
      url={https://arxiv.org/abs/2408.01605}, 
}

@misc{cwe_view_2000,
  author = {{MITRE}},
  title  = {{CWE Comprehensive View (CWE-2000)}},
  url    = {https://cwe.mitre.org/data/definitions/2000.html}
}

@misc{cwe_mapping_guidance,
  author = {{MITRE}},
  title  = {{CVE to CWE Mapping ``Root Cause Mapping'' Guidance}},
  url    = {https://cwe.mitre.org/documents/cwe_usage/guidance.html}
}

@misc{wang2024reposvulrepositorylevelhighqualityvulnerability,
      title={ReposVul: A Repository-Level High-Quality Vulnerability Dataset}, 
      author={Xinchen Wang and Ruida Hu and Cuiyun Gao and Xin-Cheng Wen and Yujia Chen and Qing Liao},
      year={2024},
      eprint={2401.13169},
      archivePrefix={arXiv},
      primaryClass={cs.CR},
      url={https://arxiv.org/abs/2401.13169}, 
}

@misc{CWE94,
  title        = {{CWE-94: Improper Control of Generation of Code ('Code Injection')}},
  URL ={https://cwe.mitre.org/data/definitions/94.html}
}

@misc{CWE79,
  title        = {{CWE-79: Improper Neutralization of Input During Web Page Generation ('Cross-site Scripting')}},
  URL = {https://cwe.mitre.org/data/definitions/79.html}
}

@misc{CWE22,
  title        = {{CWE-22: Improper Limitation of a Pathname to a Restricted Directory ('Path Traversal')}},
  URL = {https://cwe.mitre.org/data/definitions/22.html}
}

@misc{CWE89,
  title        = {{CWE-89: Improper Neutralization of Special Elements used in an SQL Command ('SQL Injection')}},
  URL = {https://cwe.mitre.org/data/definitions/89.html}
}

@misc{CWE862,
  title        = {{CWE-862: Missing Authorization}},
  URL = {https://cwe.mitre.org/data/definitions/862.html}
}

@inproceedings{SVEN,
author = {He, Jingxuan and Vechev, Martin},
title = {Large Language Models for Code: Security Hardening and Adversarial Testing},
year = {2023},
isbn = {9798400700507},
publisher = {Association for Computing Machinery},
address = {New York, NY, USA},
url = {https://doi.org/10.1145/3576915.3623175},
doi = {10.1145/3576915.3623175},
pages = {1865–1879},
numpages = {15},
location = {Copenhagen, Denmark},
series = {CCS '23}
}

@inproceedings{safecoder,
author = {He, Jingxuan and Vero, Mark and Krasnopolska, Gabriela and Vechev, Martin},
title = {Instruction tuning for secure code generation},
year = {2024},
publisher = {JMLR.org},
articleno = {723},
numpages = {20},
location = {Vienna, Austria},
series = {ICML'24}
}

@inproceedings{CoSec,
author = {Li, Dong and Yan, Meng and Zhang, Yaosheng and Liu, Zhongxin and Liu, Chao and Zhang, Xiaohong and Chen, Ting and Lo, David},
title = {CoSec: On-the-Fly Security Hardening of Code LLMs via Supervised Co-decoding},
year = {2024},
isbn = {9798400706127},
publisher = {Association for Computing Machinery},
address = {New York, NY, USA},
url = {https://doi.org/10.1145/3650212.3680371},
doi = {10.1145/3650212.3680371},
pages = {1428–1439},
numpages = {12},
location = {Vienna, Austria},
series = {ISSTA 2024}
}

@misc{SoSecure,
      title={SOSecure: Safer Code Generation with RAG and StackOverflow Discussions}, 
      author={Manisha Mukherjee and Vincent J. Hellendoorn},
      year={2026},
      eprint={2503.13654},
      archivePrefix={arXiv},
      primaryClass={cs.SE},
      url={https://arxiv.org/abs/2503.13654}, 
}

@article{promptsecure,
author = {Tony, Catherine and D\'{\i}az Ferreyra, Nicol\'{a}s E. and Mutas, Markus and Dhif, Salem and Scandariato, Riccardo},
title = {Prompting Techniques for Secure Code Generation: A Systematic Investigation},
year = {2025},
issue_date = {November 2025},
publisher = {Association for Computing Machinery},
address = {New York, NY, USA},
volume = {34},
number = {8},
issn = {1049-331X},
url = {https://doi.org/10.1145/3722108},
doi = {10.1145/3722108},
journal = {ACM Trans. Softw. Eng. Methodol.},
month = oct,
articleno = {225},
numpages = {53}
}

@misc{zhang2024seccodergeneralizablerobustsecure,
      title={SecCoder: Towards Generalizable and Robust Secure Code Generation}, 
      author={Boyu Zhang and Tianyu Du and Junkai Tong and Xuhong Zhang and Kingsum Chow and Sheng Cheng and Xun Wang and Jianwei Yin},
      year={2024},
      eprint={2410.01488},
      archivePrefix={arXiv},
      primaryClass={cs.PL},
      url={https://arxiv.org/abs/2410.01488}, 
}

@misc{codeql,
  author = {{GitHub}},
  title  = {{CodeQL}},
  url    = {https://codeql.github.com/}
}

@misc{semgrep,
  author = {{Semgrep}},
  title  = {{Semgrep App Security Platform}},
  url    = {https://semgrep.dev/}
}

@misc{weggli,
  author = {{weggli-rs}},
  title  = {{weggli: A Fast and Robust Semantic Search Tool for C and C++ Codebases}},
  url    = {https://github.com/weggli-rs/weggli}
}

@article{10.1145/3610721,
author = {Pearce, Hammond and Ahmad, Baleegh and Tan, Benjamin and Dolan-Gavitt, Brendan and Karri, Ramesh},
title = {Asleep at the Keyboard? Assessing the Security of GitHub Copilot’s Code Contributions},
year = {2025},
issue_date = {February 2025},
publisher = {Association for Computing Machinery},
address = {New York, NY, USA},
volume = {68},
number = {2},
issn = {0001-0782},
url = {https://doi.org/10.1145/3610721},
doi = {10.1145/3610721},
journal = {Commun. ACM},
month = jan,
pages = {96–105},
numpages = {10}
}

@article{10.1007/s10664-024-10590-1,
author = {Tihanyi, Norbert and Bisztray, Tamas and Ferrag, Mohamed Amine and Jain, Ridhi and Cordeiro, Lucas C.},
title = {How secure is AI-generated code: a large-scale comparison of large language models},
year = {2024},
issue_date = {Mar 2025},
publisher = {Kluwer Academic Publishers},
address = {USA},
volume = {30},
number = {2},
issn = {1382-3256},
url = {https://doi.org/10.1007/s10664-024-10590-1},
doi = {10.1007/s10664-024-10590-1},
journal = {Empirical Softw. Engg.},
month = dec,
numpages = {42}
}

@misc{qwen2_5_32b,
  author = {{Alibaba Group}},
  title  = {{Qwen2.5-Coder-32B}},
  url    = {https://huggingface.co/Qwen/Qwen2.5-Coder-32B}
}

@misc{qwen3_5_35b,
  author = {{Qwen Team}},
  title  = {{Qwen3.5-35B-A3B}},
  url    = {https://huggingface.co/Qwen/Qwen3.5-35B-A3B}
}

@misc{gemma_2025,
    title={Gemma 3},
    url={https://goo.gle/Gemma3Report},
    author={Gemma Team},
    year={2025}
}

@misc{qwen3_6_35b,
  author = {{Qwen Team}},
  title  = {{Qwen3.6-35B-A3B}},
  url    = {https://huggingface.co/Qwen/Qwen3.6-35B-A3B}
}

@misc{bge_base_en_v1_5,
  author = {{BAAI}},
  title  = {{BGE-base-en-v1.5}},
  url    = {https://huggingface.co/BAAI/bge-base-en-v1.5}
}

@misc{CWE95,
  title = {{CWE-95: Improper Neutralization of Directives in Dynamically Evaluated Code ('Eval Injection')}},
  url   = {https://cwe.mitre.org/data/definitions/95.html}
}

@misc{CWE113,
  title = {{CWE-113: Improper Neutralization of CRLF Sequences in HTTP Headers ('HTTP Request/Response Splitting')}},
  url   = {https://cwe.mitre.org/data/definitions/113.html}
}

@misc{CWE117,
  title = {{CWE-117: Improper Output Neutralization for Logs}},
  url   = {https://cwe.mitre.org/data/definitions/117.html}
}

@misc{CWE170,
  title = {{CWE-170: Improper Null Termination}},
  url   = {https://cwe.mitre.org/data/definitions/170.html}
}

@misc{CWE215,
  title = {{CWE-215: Insertion of Sensitive Information Into Debugging Code}},
  url   = {https://cwe.mitre.org/data/definitions/215.html}
}

@misc{CWE312,
  title = {{CWE-312: Cleartext Storage of Sensitive Information}},
  url   = {https://cwe.mitre.org/data/definitions/312.html}
}

@misc{CWE327,
  title = {{CWE-327: Use of a Broken or Risky Cryptographic Algorithm}},
  url   = {https://cwe.mitre.org/data/definitions/327.html}
}

@misc{CWE377,
  title = {{CWE-377: Insecure Temporary File}},
  url   = {https://cwe.mitre.org/data/definitions/377.html}
}

@misc{CWE468,
  title = {{CWE-468: Incorrect Pointer Scaling}},
  url   = {https://cwe.mitre.org/data/definitions/468.html}
}

@misc{CWE643,
  title = {{CWE-643: Improper Neutralization of Data within XPath Expressions ('XPath Injection')}},
  url   = {https://cwe.mitre.org/data/definitions/643.html}
}

@misc{CWE676,
  title = {{CWE-676: Use of Potentially Dangerous Function}},
  url   = {https://cwe.mitre.org/data/definitions/676.html}
}

@misc{CWE732,
  title = {{CWE-732: Incorrect Permission Assignment for Critical Resource}},
  url   = {https://cwe.mitre.org/data/definitions/732.html}
}

@misc{CWE776,
  title = {{CWE-776: Improper Restriction of Recursive Entity References in DTDs ('XML Entity Expansion')}},
  url   = {https://cwe.mitre.org/data/definitions/776.html}
}

@inproceedings{NEURIPS2024_9b812ee4,
 author = {Le, Hung and Zhou, Yingbo and Xiong, Caiming and Savarese, Silvio and Doyen Sahoo},
 booktitle = {Advances in Neural Information Processing Systems},
 doi = {10.52202/079017-2717},
 editor = {A. Globerson and L. Mackey and D. Belgrave and A. Fan and U. Paquet and J. Tomczak and C. Zhang},
 pages = {85546--85582},
 publisher = {Curran Associates, Inc.},
 title = {INDICT: Code Generation with Internal Dialogues of Critiques for Both Security and Helpfulness},
 url = {https://proceedings.neurips.cc/paper_files/paper/2024/file/9b812ee4b831c21e14156ced8659197c-Paper-Conference.pdf},
 volume = {37},
 year = {2024}
}

@misc{kim2024codexitysecureaiassistedcode,
      title={Codexity: Secure AI-assisted Code Generation}, 
      author={Sung Yong Kim and Zhiyu Fan and Yannic Noller and Abhik Roychoudhury},
      year={2024},
      eprint={2405.03927},
      archivePrefix={arXiv},
      primaryClass={cs.SE},
      url={https://arxiv.org/abs/2405.03927}, 
}

@inproceedings {307966,
author = {Joseph Spracklen and Raveen Wijewickrama and A H M Nazmus Sakib and Anindya Maiti and Bimal Viswanath},
title = {We Have a Package for You! A Comprehensive Analysis of Package Hallucinations by Code Generating {LLMs}},
booktitle = {34th USENIX Security Symposium (USENIX Security 25)},
year = {2025},
isbn = {978-1-939133-52-6},
address = {Seattle, WA},
pages = {3687--3706},
url = {https://www.usenix.org/conference/usenixsecurity25/presentation/spracklen},
publisher = {USENIX Association},
month = aug
}

@misc{thang2026goodvibesecuritybyvibellmbasedcode,
      title={GoodVibe: Security-by-Vibe for LLM-Based Code Generation}, 
      author={Maximilian Thang and Lichao Wu and Sasha Behrouzi and Mohamadreza Rostami and Jona te Lintelo and Stjepan Picek and Ahmad-Reza Sadeghi},
      year={2026},
      eprint={2602.10778},
      archivePrefix={arXiv},
      primaryClass={cs.CR},
      url={https://arxiv.org/abs/2602.10778}, 
}

@inproceedings{parvez-etal-2021-retrieval-augmented,
    title = "Retrieval Augmented Code Generation and Summarization",
    author = "Parvez, Md Rizwan  and
      Ahmad, Wasi  and
      Chakraborty, Saikat  and
      Ray, Baishakhi  and
      Chang, Kai-Wei",
    editor = "Moens, Marie-Francine  and
      Huang, Xuanjing  and
      Specia, Lucia  and
      Yih, Scott Wen-tau",
    booktitle = "Findings of the Association for Computational Linguistics: EMNLP 2021",
    month = nov,
    year = "2021",
    address = "Punta Cana, Dominican Republic",
    publisher = "Association for Computational Linguistics",
    url = "https://aclanthology.org/2021.findings-emnlp.232/",
    doi = "10.18653/v1/2021.findings-emnlp.232",
    pages = "2719--2734"
}

@misc{zhou2023docpromptinggeneratingcoderetrieving,
      title={DocPrompting: Generating Code by Retrieving the Docs}, 
      author={Shuyan Zhou and Uri Alon and Frank F. Xu and Zhiruo Wang and Zhengbao Jiang and Graham Neubig},
      year={2023},
      eprint={2207.05987},
      archivePrefix={arXiv},
      primaryClass={cs.CL},
      url={https://arxiv.org/abs/2207.05987}, 
}

@inproceedings{zhang-etal-2023-repocoder,
    title = "{R}epo{C}oder: Repository-Level Code Completion Through Iterative Retrieval and Generation",
    author = "Zhang, Fengji  and
      Chen, Bei  and
      Zhang, Yue  and
      Keung, Jacky  and
      Liu, Jin  and
      Zan, Daoguang  and
      Mao, Yi  and
      Lou, Jian-Guang  and
      Chen, Weizhu",
    editor = "Bouamor, Houda  and
      Pino, Juan  and
      Bali, Kalika",
    booktitle = "Proceedings of the 2023 Conference on Empirical Methods in Natural Language Processing",
    month = dec,
    year = "2023",
    address = "Singapore",
    publisher = "Association for Computational Linguistics",
    url = "https://aclanthology.org/2023.emnlp-main.151/",
    doi = "10.18653/v1/2023.emnlp-main.151",
    pages = "2471--2484"
}

@article{Tihanyi2024,
  author  = {Tihanyi, Norbert and Bisztray, Tamas and Ferrag, Mohamed Amine and Jain, Ridhi and Cordeiro, Lucas C.},
  title   = {{How secure is AI-generated code: a large-scale comparison of large language models}},
  journal = {Empirical Software Engineering},
  year    = {2024},
  volume  = {30},
  number  = {2},
  pages   = {47},
  doi     = {10.1007/s10664-024-10590-1},
  url     = {https://doi.org/10.1007/s10664-024-10590-1}
}

@misc{kharma2026securityqualityllmgeneratedcode,
      title={Security and Quality in LLM-Generated Code: A Multi-Language, Multi-Model Analysis}, 
      author={Mohammed Kharma and Soohyeon Choi and Mohammed AlKhanafseh and David Mohaisen},
      year={2026},
      eprint={2502.01853},
      archivePrefix={arXiv},
      primaryClass={cs.CR},
      url={https://arxiv.org/abs/2502.01853}, 
}

@article{10.1145/3797277,
author = {Du, Xueying and Zheng, Geng and Wang, Kaixin and Zou, Yi and Wang, Yujia and Deng, Wentai and Feng, Jiayi and Liu, Mingwei and Chen, Bihuan and Peng, Xin and Ma, Tao and Lou, Yiling},
title = {Vul-RAG: Enhancing LLM-based Vulnerability Detection via Knowledge-level RAG},
year = {2026},
publisher = {Association for Computing Machinery},
address = {New York, NY, USA},
issn = {1049-331X},
url = {https://doi.org/10.1145/3797277},
doi = {10.1145/3797277},
note = {Just Accepted},
journal = {ACM Trans. Softw. Eng. Methodol.},
month = feb
}

@article{empirical_Study,
author = {Yang, Zezhou and Chen, Sirong and Gao, Cuiyun and Li, Zhenhao and Hu, Xing and Liu, Kui and Xia, Xin},
title = {An Empirical Study of Retrieval-Augmented Code Generation: Challenges and Opportunities},
year = {2025},
issue_date = {September 2025},
publisher = {Association for Computing Machinery},
address = {New York, NY, USA},
volume = {34},
number = {7},
issn = {1049-331X},
url = {https://doi.org/10.1145/3717061},
doi = {10.1145/3717061},
journal = {ACM Trans. Softw. Eng. Methodol.},
month = aug,
articleno = {188},
numpages = {28}
}

@inproceedings{llm_control_code,
author = {Koziolek, Heiko and Gr\"{u}ner, Sten and Hark, Rhaban and Ashiwal, Virendra and Linsbauer, Sofia and Eskandani, Nafise},
title = {LLM-based and Retrieval-Augmented Control Code Generation},
year = {2024},
isbn = {9798400705793},
publisher = {Association for Computing Machinery},
address = {New York, NY, USA},
url = {https://doi.org/10.1145/3643795.3648384},
doi = {10.1145/3643795.3648384},
booktitle = {Proceedings of the 1st International Workshop on Large Language Models for Code},
pages = {22–29},
numpages = {8},
location = {Lisbon, Portugal},
series = {LLM4Code '24}
}

@inproceedings{lu-etal-2022-reacc,
    title = "{R}e{ACC}: A Retrieval-Augmented Code Completion Framework",
    author = "Lu, Shuai  and
      Duan, Nan  and
      Han, Hojae  and
      Guo, Daya  and
      Hwang, Seung-won  and
      Svyatkovskiy, Alexey",
    editor = "Muresan, Smaranda  and
      Nakov, Preslav  and
      Villavicencio, Aline",
    booktitle = "Proceedings of the 60th Annual Meeting of the Association for Computational Linguistics (Volume 1: Long Papers)",
    month = may,
    year = "2022",
    address = "Dublin, Ireland",
    publisher = "Association for Computational Linguistics",
    url = "https://aclanthology.org/2022.acl-long.431/",
    doi = "10.18653/v1/2022.acl-long.431",
    pages = "6227--6240"
}

\section*{Appendix}
\subsection{LLM Curator Prompt}
\label{app:curator-prompt}

\clearpage

\begin{figure*}[t]
\centering
\begin{tcolorbox}[
    colback=gray!5!white,
    colframe=black,
    width=\dimexpr\textwidth-0.30in\relax,
    boxrule=0.5pt,
    arc=4pt,
    outer arc=4pt,
    left=2mm,
    right=2mm,
    top=1mm,
    bottom=1mm
]

\textbf{LLM Curator Prompt Template (Part 1):}

\begin{lstlisting}[
    basicstyle=\ttfamily\scriptsize,
    breaklines=true,
    columns=fullflexible,
    keepspaces=true,
    showstringspaces=false,
    aboveskip=2pt,
    belowskip=0pt
]
You are a software security expert building a knowledge base of
vulnerability-fix pairs for secure code generation. You are given
a real vulnerability: a function before a security fix, the same
function after the fix, and metadata describing the flaw.

Extract two descriptions and two code snippets that will later be
used to retrieve and augment security knowledge for new code
generation tasks.

Describe only what is evident in the provided code and metadata.
The function shown may not be the complete context. Do not infer or
invent surrounding code, callers, or behavior you cannot see.

The root_cause, fixing_pattern, and fixed_snippet are shown together
to the code generator, so they must form a coherent set:
- root_cause names the flaw;
- fixing_pattern states the transferable rule that resolves it;
- fixed_snippet shows that rule in code.

Produce the following fields:

1. root_cause:
The specific coding mistake in this function that introduces the
vulnerability. Keep it self-contained, concrete, and instance-specific.
Do not restate the generic CWE definition.

2. fixing_pattern:
The remediation expressed as a transferable principle that can be
applied to new code. Prefer the general security rule over a
description of the specific edit.

3. vulnerable_snippet:
A short retrieval target centered on the vulnerable pattern.
Keep only the lines needed to expose the security-relevant data flow
or dangerous operation. Long literals or embedded query text may be
replaced by a placeholder. Otherwise preserve the original structure.
The snippet need not be complete or runnable.

4. fixed_snippet:
A concise secure idiom copied verbatim from the post-fix version.
Keep only the statements that constitute the fix and the minimum
surrounding lines needed to understand it. Do not rewrite, rename,
reformat, or add "..." padding.
\end{lstlisting}

\end{tcolorbox}
\caption{Prompt template for the LLM Curator: facet derivation.}
\label{fig:curator-prompt-part1}
\end{figure*}

\clearpage

\begin{figure*}[t]
\centering
\begin{tcolorbox}[
    colback=gray!5!white,
    colframe=black,
    width=\dimexpr\textwidth-0.30in\relax,
    boxrule=0.5pt,
    arc=4pt,
    outer arc=4pt,
    left=2mm,
    right=2mm,
    top=1mm,
    bottom=1mm
]

\textbf{LLM Curator Prompt Template (Part 2):}

\begin{lstlisting}[
    basicstyle=\ttfamily\scriptsize,
    breaklines=true,
    columns=fullflexible,
    keepspaces=true,
    showstringspaces=false,
    aboveskip=2pt,
    belowskip=0pt
]
Rules:
- root_cause must describe the flaw visible in vulnerable_snippet.
- fixing_pattern and fixed_snippet must describe and show the same fix.
- Do not repeat the same information across fields.
- If the vulnerability or fix cannot be reliably determined, set
  "refused" to true and do not invent missing context.
- Otherwise set "refused" to false.
- Grade the pair using "quality":
    clean: one clearly visible and transferable security fix.
    mixed: the security fix is identifiable but entangled or partial.
    weak: the actual fix is absent, mismatched, or not transferable.

Input:

Function Before the Fix:
<function_before>

Function After the Fix:
<function_after>

CVE Description:
<cve_description>

Commit Message:
<commit_message>

Connected CWEs:
<cwe_ids_and_names>

Return JSON only with:

{
  "root_cause": "...",
  "fixing_pattern": "...",
  "vulnerable_snippet": "...",
  "fixed_snippet": "...",
  "quality": "clean | mixed | weak",
  "refused": true | false
}
\end{lstlisting}

\end{tcolorbox}
\caption{Prompt template for the LLM Curator: curation rules, inputs, and output format.}
\label{fig:curator-prompt-part2}
\end{figure*}

\begin{figure*}[t]
\centering
\begin{tcolorbox}[
    colback=gray!5!white,
    colframe=black,
    width=\dimexpr\textwidth-0.30in\relax,
    boxrule=0.5pt,
    arc=4pt,
    outer arc=4pt,
    left=2mm,
    right=2mm,
    top=1mm,
    bottom=1mm
]

\textbf{Planner Agent Prompt Template:}

\begin{lstlisting}[
    basicstyle=\ttfamily\scriptsize,
    breaklines=true,
    columns=fullflexible,
    keepspaces=true,
    showstringspaces=false,
    aboveskip=2pt,
    belowskip=0pt
]
You are a retrieval planning component in a secure code generation pipeline. You are given a coding task to identify how the code could be insecure.

A task may be insecure in several genuinely distinct ways: fundamentally different flaws with different root causes. Predict several only when they are truly distinct. Do not add a weakness that is a broader or narrower version of one you already named; name only the one that best fits. If you are unsure if any weakness is present, do not predict any. Do not speculate based on how the code might be modified later.

First state the task's language, exactly one of:
"Java", "Python", "C", "C++".

For each distinct weakness, produce an entry with three fields:

1. cwe_id:
The integer ID of the specific weakness, e.g., 22 for path traversal. Predict only current, non-deprecated CWEs. Never predict a CWE category; always select the concrete weakness rather than a broader class.

2. root_cause:
The specific reason THIS task could exhibit the weakness. Keep it concrete and instance-specific; do not restate the generic CWE definition.

Good:
"A file name taken from the request URL is joined to the base directory without normalization, so parent-directory segments could escape the intended directory."

Bad:
"Improper limitation of a pathname to a restricted directory."
(generic CWE boilerplate)

3. code_snippet:
A short fragment in the task's language showing the shape of the flaw. Keep it to a few legible lines that convey the vulnerable pattern. It need not be complete, runnable, or syntactically valid. Each weakness gets its own snippet.

Input:
<coding_task>

Respond with a JSON object with exactly this shape:

{
  "language": "Python",
  "reasoning": "...",
  "weaknesses": [
    {
      "cwe_id": 22,
      "root_cause": "...",
      "code_snippet": "..."
    }
  ]
}

Use these exact field names. Respond with JSON only. Do not include text, markdown, or code fences outside the JSON.
\end{lstlisting}

\end{tcolorbox}
\caption{Prompt template for the Planner Agent.}
\label{fig:planner-prompt}
\end{figure*}

\clearpage

\begin{figure*}[t]
\centering
\begin{tcolorbox}[
    colback=gray!5!white,
    colframe=black,
    width=\dimexpr\textwidth-0.30in\relax,
    boxrule=0.5pt,
    arc=4pt,
    outer arc=4pt,
    left=2mm,
    right=2mm,
    top=1mm,
    bottom=1mm
]

\textbf{Selection Agent Prompt Template (Part 1):}

\begin{lstlisting}[
    basicstyle=\ttfamily\scriptsize,
    breaklines=true,
    columns=fullflexible,
    keepspaces=true,
    showstringspaces=false,
    aboveskip=2pt,
    belowskip=0pt
]
You are a retrieval selector for a secure code generation pipeline.

You receive:
- one coding task,
- the pipeline's prediction of the relevant security weakness, including a reasoning paragraph and retrieval units with CWE IDs and root-cause descriptions,
- a set of candidate examples retrieved from a vulnerability-fix knowledge base.

The task is a code-completion task. The provided code is an unfinished stub, and the generator will complete it. Judge the predicted weakness against the task's intent.

You make two kinds of judgment.

FIRST, judge whether the prediction is sound.

A prediction is SOUND when its reasoning and retrieval units plausibly describe a weakness that the completed code is likely to exhibit given the requested functionality. Treat soundness as the default unless the prediction clearly does not fit the task.

A prediction is UNSOUND when the predicted weakness does not match the task's intent, input source, or operation. The prediction is not unsound merely because the vulnerable code is not yet present in the unfinished stub.

If the prediction is SOUND:
- set "prediction_is_sound" to true,
- set "feedback" to null,
- evaluate every retrieved candidate and return the results in "verdicts".

If the prediction is UNSOUND:
- set "prediction_is_sound" to false,
- set "verdicts" to null,
- provide concise feedback explaining why the prediction is incorrect,
- identify the CWE or weakness that better matches the task,
- point to the relevant part of the task.

Do not filter candidates when the prediction is unsound.
\end{lstlisting}

\end{tcolorbox}
\caption{Prompt template for the Selection Agent: prediction validation and feedback.}
\label{fig:selection-prompt-part1}
\end{figure*}

\clearpage

\begin{figure*}[t]
\centering
\begin{tcolorbox}[
    colback=gray!5!white,
    colframe=black,
    width=\dimexpr\textwidth-0.30in\relax,
    boxrule=0.5pt,
    arc=4pt,
    outer arc=4pt,
    left=2mm,
    right=2mm,
    top=1mm,
    bottom=1mm
]

\textbf{Selection Agent Prompt Template (Part 2):}

\begin{lstlisting}[
    basicstyle=\ttfamily\scriptsize,
    breaklines=true,
    columns=fullflexible,
    keepspaces=true,
    showstringspaces=false,
    aboveskip=2pt,
    belowskip=0pt
]
SECOND, when the prediction is sound, filter the candidates.

Keep a candidate when its security lesson addresses the same or a closely related weakness relevant to the task. The surrounding functionality, code structure, function names, or API do not need to match.

Candidates from another programming language may be kept when the underlying fixing principle transfers to the task language.

Drop a candidate only when its security lesson targets a different weakness or could mislead the generator toward an inappropriate change. When uncertain whether a candidate is useful, keep it.

Input:

Coding Task:
<coding_task>

Planner Reasoning:
<planner_reasoning>

Retrieval Units:
<cwe_id, root_cause, ...>

Retrieved Candidates:
<candidate_0>
<candidate_1>
...

If the prediction is sound, return:

{
  "prediction_is_sound": true,
  "feedback": null,
  "verdicts": [
    {
      "index": 0,
      "reason": "...",
      "keep": true
    }
  ]
}

If the prediction is unsound, return:

{
  "prediction_is_sound": false,
  "feedback": "...",
  "verdicts": null
}

Include one verdict for every candidate when the prediction is sound. Use the exact field names above. Respond with JSON only. Do not include text, markdown, or code fences outside the JSON.
\end{lstlisting}

\end{tcolorbox}
\caption{Prompt template for the Selection Agent: candidate filtering and output format.}
\label{fig:selection-prompt-part2}
\end{figure*}

\end{document}